\documentclass[aps,twocolumn,pra,showkeys,superscriptaddress]{revtex4-2}
\usepackage{mathrsfs}
\usepackage{array}
\usepackage{booktabs}
\usepackage{amsmath}
\usepackage{amsfonts}
\usepackage{amssymb}
\usepackage{graphicx,color}
\usepackage[colorlinks={true}]{hyperref}
\hypersetup{colorlinks=true,linkcolor=red,citecolor=blue,urlcolor=blue}
\usepackage{bm}
\usepackage{orcidlink}
\usepackage{comment}
\begin{document}

% \title{One- and Two-Dimensional Electromagnetically Induced Gratings through Quantum Qubit System via Weak Decay Field}
\title{Reservoir-controlled electromagnetically induced gratings in a weakly driven two-level medium}

\author{Amjad Hussain}
\affiliation{Department of Physics, Abbottabad University of Science and Technology, Havellian, 22500, Pakistan}

\author{Hamid Arian Zad~\!\!\orcidlink{0000-0002-1348-1777}}
\email{hamid.arian.zad@upjs.sk}
\address{Department of Theoretical Physics and Astrophysics, Faculty of Science of P. J. \v{S}af{\'a}rik University, Park Angelinum 9, 040 01 Ko\v{s}ice, Slovak Republic}

\author{Amjad Sohail~\!\!\orcidlink{0000-0001-8777-7928}}
\affiliation{Department of Physics, Government College University, Allama Iqbal Road, 38000 Faisalabad, Pakistan}

\author{Hazrat Ali~\!\!\orcidlink{0000-0003-1957-3629}} \email{yamanuom@gmail.com}
\affiliation{Department of Physics, Abbottabad University of Science and Technology, Havellian, 22500, Pakistan}

\author{Michal Ja{\v s}{\v c}ur~\!\!\orcidlink{0000-0003-0826-1961}}%
\address{Department of Theoretical Physics and Astrophysics, Faculty of Science of P. J. \v{S}af{\'a}rik University, Park Angelinum 9, 040 01 Ko\v{s}ice, Slovak Republic}%

\author{Saeed Haddadi~\!\!\orcidlink{0000-0002-1596-0763}} \email{haddadi@ipm.ir}
\affiliation{School of Particles and Accelerators, Institute for Research in Fundamental Sciences (IPM), P.O. Box 19395-5531, Tehran, Iran}

\begin{abstract}
We theoretically investigate the transmission and diffraction of a weak probe field from an electromagnetically induced grating formed in a weakly driven two-level medium coupled to engineered quantum reservoirs. Using a perturbative solution of the optical Bloch equations in the weak-driving regime, we analyze how normal-vacuum, thermal, and broadband squeezed-vacuum environments modify the probe susceptibility and consequently reshape both the spatial transmission function and the far-field diffraction patterns. We show that reservoir statistics have a pronounced impact on the diffraction response by altering the amplitude and phase of the induced grating. Thermal reservoirs enhance the transmission modulation and increase the intensity of the dominant diffraction orders, whereas squeezed-vacuum reservoirs generate strongly phase-sensitive modifications that selectively redistribute optical power among diffraction channels. We further demonstrate that the detuning between the squeezed reservoir and the driving field provides an efficient mechanism for controlling diffraction directionality, leading to substantial amplification of selected angular orders. In two-dimensional geometries, squeezed-vacuum correlations produce highly structured phase landscapes and strongly anisotropic diffraction patterns, enabling directional enhancement of specific diffraction channels while suppressing others. These results establish reservoir engineering as a versatile approach for controlling transmission, diffraction efficiency, and angular selectivity in minimal two-level systems, with potential applications in programmable photonic devices, beam steering, and quantum optical platforms.

\end{abstract}

%\date{\today}

\maketitle

\section{Introduction} 

The interaction of light with atomic systems gives rise to a variety of remarkable quantum-optical phenomena, among which electromagnetically induced transparency (EIT) has attracted sustained attention \cite{scully2012quantum,gerry2012introductory}. Originating from quantum coherence and interference effects \cite{stern2013nanoscale,harris1997eit,fleischhauer2002memory}, EIT enables an otherwise absorptive medium to become transparent to a weak probe field \cite{fleischhauer2005review}. Based on this mechanism, the concept of electromagnetically induced gratings (EIG) has emerged \cite{mitsunaga1999eig,deng2003ultraslow}, where standing-wave control fields generate spatial modulations in the absorptive and refractive properties of atomic media \cite{wen2011talbot,ba2012magnetic}.

EIG systems have demonstrated considerable potential for applications including all-optical switching and optical diode action \cite{deng2003ultraslow,brown2005switching}, optical bistability \cite{zhai2001bistability}, and the realization of photonic band-gap structures \cite{chen2017router}. Beyond conventional one-dimensional configurations, G. L. Cheng \textit{et al.} \cite{cheng2016JPhysB} proposed a two-dimensional gain-phase grating in a weakly driven two-level atomic system and showed that orthogonal standing-wave fields can generate pronounced high-order diffraction with controllable diffraction efficiencies. Later, B. Wang \textit{et al.} \cite{wang2022rydberg} investigated a two-dimensional asymmetric EIG in Rydberg atoms and demonstrated that the diffraction pattern can be tailored to exhibit highly asymmetric intensity distributions through appropriate modulation of the control-field parameters. In another work \cite{asadpour2021azimuthal} an azimuthally modulated EIGs based on structured light was introduced that provides an additional degree of control over diffraction via the spatial phase and orbital structure of the driving fields.

 Experimentally, thermal atomic vapors remain among the most widely used platforms for realizing EIGs \cite{deng2001rubidium,kang2003efficient,yan2003highorder}, with diffraction and tunability extensively explored in rubidium-vapor systems \cite{yan2001switching,zhang2018talbot,yan2003highorder}. These developments have significantly broadened the scope of EIG research, particularly in multidimensional configurations \cite{wang2022rydberg,asadpour2021azimuthal}, which offer enhanced flexibility for diffraction engineering, light manipulation, and the exploration of non-Hermitian optical phenomena.
At the same time, squeezed vacuum fields as the nonclassical states of light characterized by reduced quantum fluctuations in one field quadrature have attracted considerable attention because of their ability to enhance quantum sensing and precision measurement protocols \cite{xu2019sensing,grace2020gyroscope,malitesta2023distributed,zhang2022quadratic,caves1981noise,vahlbruch2016fifteen,braunstein2005continuous,gisin2007communication}. Such quantum states now play an essential role in areas including quantum communication \cite{gisin2007communication,schnabel2017squeezed}, gravitational-wave detection \cite{schnabel2017squeezed,loudon2000light}, and nonlinear quantum optics \cite{walls2008quantum,walls2008quantum2}. 
In particular, optical parametric oscillators  \cite{wu1986parametric,vahlbruch2016fifteen,serikawa2017broadband,chelkowski2024coherent} can generate broadband squeezed vacuum capable of suppressing vacuum fluctuations significantly below the standard quantum limit \cite{anderson1995quadrature,silberhorn2001epr}. The properties of broadband (multimode) squeezed radiation are commonly characterized by two key quantities including the average photon number spectrum $n(\omega_0)$ and the two-mode correlation function $p(\omega_0)$, which quantifies squeezing-induced correlations between frequency-symmetric modes around a central frequency $\omega_0$ \cite{caves1985formalism1,schumaker1985formalism2}. These balanced intermode correlations, centered around the probe frequency, distinguish broadband squeezed radiation from an ensemble of independent squeezed modes. Experimental observations have confirmed the importance of such correlations in determining the interaction between atoms and squeezed electromagnetic environments \cite{lu1998interference,christ2011multimode,roslund2014quantum}.

In describing the interaction of atoms with squeezed radiation, it is often assumed that the squeezed-vacuum bandwidth is much larger than the atomic linewidth $\gamma$. Under this condition, the squeezed modes may be treated as effectively delta-correlated in time, thereby behaving as squeezed quantum noise \cite{gardiner2004noise,collett1984squeezing}. This approximation enables the derivation of reservoir-induced damping terms involving the squeezing parameters $N$ and $M$ within both master-equation \cite{gardiner1986inhibition} and Heisenberg-picture approaches. However, several experimental implementations have realized squeezed-vacuum sources with bandwidths comparable to $\gamma$ \cite{wu1986parametric,polzik1992spectroscopy}, leading to the regime of finite-bandwidth squeezed vacuum radiation. In this regime, the atom–field interaction becomes considerably richer, producing modified resonance fluorescence and nontrivial spectral effects observed in both numerical \cite{zhu1990fluorescence,jin1990bandwidth} and analytical studies \cite{hassan2013nonlinear,lu1998frequency,parkins1993twophoton}. These effects arise from the frequency dependence of the squeezing spectrum and its associated correlation amplitudes.

Motivated by these developments, the present work investigates the formation and control of EIGs in a weakly driven two-level atomic system coupled to structured reservoirs. In contrast to standard EIG schemes based on $\Lambda$- or ladder-type EIT media, where diffraction control is primarily governed by multilevel quantum interference, our model isolates how reservoir statistics alone reshape the susceptibility and diffraction response of a minimal two-level medium. Specifically, we examine how thermal and broadband squeezed-vacuum reservoirs modify the transmission and diffraction properties of a weak probe field, with particular emphasis on the roles played by reservoir occupation, squeezing-induced correlations, reservoir detuning, and the decay of the driving field.

The squeezed reservoir is modeled as a broadband squeezed vacuum characterized by the photon population $N$ and the two-mode correlation parameter $M$. Such squeezed environments are experimentally accessible in both optical and microwave quantum platforms. A strong pump laser drives a second-order nonlinear medium ($\chi^{(2)}$), producing correlated photon pairs around a central frequency $\omega_{\mathrm p}$. When the cavity bandwidth of the optical parametric oscillators exceeds the atomic linewidth, the emitted field behaves effectively as a broadband squeezed reservoir interacting with the atomic transition. In superconducting circuit architectures, an analogous squeezed reservoir can be engineered using Josephson parametric amplifiers, Josephson traveling-wave parametric amplifiers, or related microwave parametric devices \cite{kaertner1990squeezed,qiu2022jtwpa,haider2019jtwpa}. These systems generate broadband squeezed microwave fields through nonlinear Josephson interactions and can be coupled directly to superconducting qubits or transmission-line resonators. Unlike natural atoms, superconducting artificial atoms provide an exceptionally clean realization of effective two-level systems, making them particularly attractive for investigating reservoir-engineered dissipation and squeezed-bath effects. Within both optical and superconducting implementations, the reservoir detuning parameter $\delta$ which plays an important role in the present analysis can be experimentally controlled by tuning the central frequency of the squeezed source relative to the coherent driving field frequency $\omega_{\mathrm 0}$. In optical systems, this may be achieved through adjustments of the pump frequency or cavity resonance of the optical parametric oscillators, while in superconducting circuits, it can be controlled electronically through the pump and bias conditions of the parametric amplifier. Consequently, the detuned squeezed-vacuum regime investigated here is practically accessible and provides a route for controlling diffraction directionality and transmission enhancement via reservoir engineering.

Finally, we note that the thermal reservoir considered in this work should be interpreted as an effective incoherent broadband environment. In optical-frequency implementations, equilibrium thermal photon populations are typically negligible at room temperature. Nevertheless, similar incoherent reservoirs can be synthesized using broadband noisy injection fields. In microwave-frequency superconducting platforms, finite thermal occupations corresponding to the parameter regime explored here are  realistic under cryogenic conditions \cite{corcoles2011protecting,krinner2019cryogenic,epjqt2025cryogenic}.

The paper is structured as follows. Section~\ref{sec:model} introduces the theoretical model for the driven two-level system coupled to structured reservoirs and derives the Bloch equations governing the atomic dynamics. Analytical expressions for the probe transmission and susceptibility are obtained in the weak-driving regime. Section~\ref{sec:results} presents numerical results for both one- and two-dimensional electromagnetically induced gratings, and demonstrates how reservoir statistics together with the control parameters influence the spatial transmission profiles and far-field diffraction patterns. Particular attention is paid to the phase-sensitive correlations of the squeezed reservoir and their ability to selectively enhance specific diffraction orders. Finally, Section~\ref{sec:conclusions} summarizes the main findings and discusses implications for reservoir-engineered quantum-optical devices, including beam steering, angular filtering, and programmable photonic routing.

\section{Theoretical Model}
\label{sec:model}

\begin{figure*}
	\centering
	\includegraphics[scale=0.28]{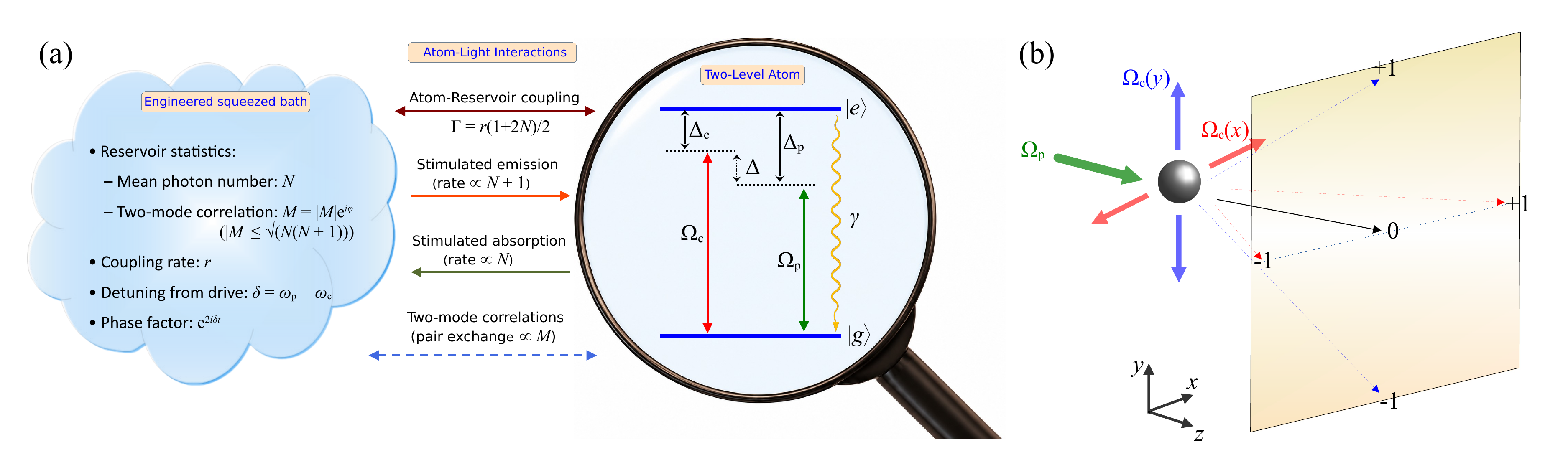}
	\vspace{-0.5cm}
	\caption{
		(a) Effective two-level atomic system driven by a standing-wave control field with Rabi frequency $\Omega_{\text{c}}$ and probed by a weak travelling field with Rabi frequency $\Omega_{\text{p}}$. The control-field detuning is $\Delta_{\text{c}}$, while the probe detuning relative to the control field is $\Delta_{\text{p}}$. The atom interacts with an engineered broadband reservoir characterized by mean photon number $N$, squeezing correlation $M$, and detuning $\delta$. The reservoir coupling gives rise to the effective coherence decay rate $\Gamma$, while $\gamma$ denotes spontaneous emission. (b) Spatial geometry of the induced grating. The weak probe field propagates along $z$ through a thin medium whose susceptibility is modulated by standing-wave control fields along the transverse directions. This produces a spatial transmission function $T(x)$ in one dimension or $T(x,y)$ in two dimensions, leading to far-field diffraction patterns $I_{\text{p}}(\theta)$ and $I_{\text{p}}(\theta_x,\theta_y)$, respectively.
	}
	\label{fig:system}
\end{figure*}

We consider an ensemble of identical effective two-level atoms with ground state $|g\rangle$ and excited state $|e\rangle$ separated by transition frequency $\omega_0$. The atoms are driven by a standing-wave control field and interrogated by a weak probe field while simultaneously interacting with an engineered broadband electromagnetic reservoir. Depending on its statistics, the reservoir may correspond to normal vacuum (NV), an effective thermal field (TF), or broadband squeezed vacuum (SV). The physical configuration is illustrated schematically in Fig.~\ref{fig:system}.
The purpose of the model is not to reproduce the complete internal structure of a specific multilevel atom, but rather to isolate the minimal mechanism through which reservoir statistics modify the optical susceptibility and consequently reshape the transmission and diffraction properties of an EIG. The system should therefore be viewed as an effective two-level transition, realizable either through a nearly closed optical transition in atomic media or through artificial atoms such as superconducting qubits.

\subsection{Microscopic Hamiltonian}

The total Hamiltonian of the system is
\begin{equation}
	\label{Eq:TotalHamiltonian}
	H = H_{\rm A}+H_{\rm F}+H_{\rm R}+H_{\rm AF}+H_{\rm AR},
\end{equation}
where $H_{\rm A}= \hbar\omega_0\hat{\sigma}_{+}\hat{\sigma}_{-}$ describes the free atomic transition, with $\hat{\sigma}_{\pm}$ the atomic raising and lowering operators. The terms $H_{\rm F}$ and $H_{\rm R}$ represent the free coherent fields and the engineered reservoir, while $H_{\rm AF}$ and $H_{\rm AR}$ describe the atom--field and atom--reservoir interactions, respectively.
The total applied electric field is $\mathbf{E}(\mathbf{r},t)= \mathbf{E}_{\text{c}}(\mathbf{r},t)+\mathbf{E}_{\text{p}}(\mathbf{r},t)$, with $\mathbf{r}=(x,y,z)$ being the position vector and the time $t$. The standing-wave control field and the weak travelling probe field are taken as
\begin{subequations}\label{Eq:SWFields}
	\begin{align}
		\mathbf{E}_{\text{c}}(\mathbf{r},t)
		&=
		\mathbf{e}_{\text{c}}E_{\text{c}0}
		\sin(k_{\text{c}}x)\cos(\omega_{\text{c}}t),
		\label{Eq:SW_c}
		\\
		\mathbf{E}_{\text{p}}(\mathbf{r},t)
		&=
		\mathbf{e}_{\text{p}}E_{\text{p}0}
		e^{i(k_{\text{p}}z-\omega_{\text{p}}t)}
		+\mathrm{c.c.},
		\label{Eq:SW_p}
	\end{align}
\end{subequations}
where $\mathbf{e}_{\text{c}}$ and $\mathbf{e}_{\text{p}}$ are polarization unit vectors, $E_{\text{c}0}$ and $E_{\text{p}0}$ are the field amplitudes, $\omega_{\text{c}}$ and $\omega_{\text{p}}$ are the control- and probe-field frequencies, and $k_{\text{c}}=2\pi/\lambda_{\text{c}}$ and $k_{\text{p}}=2\pi/\lambda_{\text{p}}$ are the corresponding wave numbers. The standing-wave modulation is along the transverse $x$ direction, whereas the probe field propagates along the longitudinal $z$ direction.

Within the electric-dipole approximation, the atom--field interaction is $H_{\rm AF}=-\mathbf{d}\cdot\mathbf{E}(\mathbf{r},t)$, where $\mathbf{d}=\boldsymbol{\mu}(\hat{\sigma}_{+}+\hat{\sigma}_{-})$ is defined as the dipole operator and $\boldsymbol{\mu}$ indicates the transition dipole moment. After applying the rotating-wave approximation, the interaction Hamiltonian becomes
\begin{equation}
	H_{\rm AF}= -\hbar
	\Big[\Omega_{\text{c}}(x)e^{-i\omega_{\text{c}}t}
	+\Omega_{\text{p}}e^{-i\omega_{\text{p}}t}
	\Big]\hat{\sigma}_{+}
	+{\rm H.c.},
\end{equation}
with  $\Omega_{\text{c}}(x)=\Omega_{\text{c}0}\sin(k_{\text{c}}x)$ being the position-dependent control-field Rabi frequency, in which  $\Omega_{\text{c}0}=(\boldsymbol{\mu}\cdot\mathbf{e}_{\text{c}}E_{\text{c}0})/\hbar$ denotes peak amplitude, and $\Omega_{\text{p}}=(\boldsymbol{\mu}\cdot\mathbf{e}_{\text{p}}E_{\text{p}0})/\hbar$ is the probe-field Rabi frequency. Throughout this work the probe field is assumed considerably weak ($\Omega_{\text{p}}\ll\Omega_{\text{c}0}$), so that the medium response remains linear in the probe field.

The broadband reservoir is characterized by a continuum of bosonic modes, $H_{\rm R}= \sum_{\lambda} \hbar \omega_{\lambda} b_{\lambda}^{\dagger} b_{\lambda}$, where $b_\lambda^{\dagger}$ and $b_\lambda$ are creation and annihilation operators for reservoir photons. The atom--reservoir coupling is consequently described by
\begin{equation}
	H_{\rm AR}= \hbar \sum_{\lambda} \left( g_{\lambda}\hat{\sigma}_{+}\hat{b}_{\lambda} + g_{\lambda}^{*}\hat{\sigma}_{-}\hat{b}_{\lambda}^{\dagger} \right).
\end{equation}
For a squeezed reservoir, the field correlations satisfy
\begin{subequations}\label{eq:squeezed_corr}
	\begin{align}
		\langle \hat{b}_{\lambda}^{\dagger}\hat{b}_{\lambda'}\rangle
		&=
		N \delta_{\lambda\lambda'},
		\label{eq:squeezed_corr_a}
		\\
		\langle \hat{b}_{\lambda}\hat{b}_{\lambda'}^{\dagger}\rangle
		&=
		(N+1)\delta_{\lambda\lambda'},
		\label{eq:squeezed_corr_b}
		\\
		\langle \hat{b}_{\lambda}\hat{b}_{\lambda'}\rangle
		&=
		M e^{-2i\omega_\text{r} t}
		\delta_{\lambda+\lambda',\,2\lambda_\text{r}},
		\label{eq:squeezed_corr_c}
	\end{align}
\end{subequations}
where $\lambda$ and $\lambda'$ are mode indices, $\lambda_\text{r}$ is the central mode index of the squeezed reservoir such that $\omega_{\lambda_\text{r}}=\omega_\text{r}$ where $\omega_\text{r}$ is the central frequency of the broadband reservoir. $N$ is the mean reservoir photon number, and $M=|M|e^{i\phi}$ characterizes the phase-sensitive two-mode squeezing correlations with the squeezing angle $\phi$ in the phase space. We set $\phi = 0$ to align the squeezed quadrature with the atomic coherence excited by the control field, leading to minimizing the effective decay rate $\Gamma_{\text{eff}} = \Gamma - |M|$ where $\Gamma=\mathcal{R}(1+2N)/2$ and $\mathcal{R}$ denotes the atom-reservoir coupling strength, and maximizing both the spatial modulation depth of the susceptibility and the resulting diffraction efficiency. The Kronecker delta $\delta_{\lambda+\lambda',\,2\lambda_\text{r}}$ enforces pairing of modes symmetrically around $\lambda_\text{r}$, i.e., $\omega_{\lambda}+\omega_{\lambda'} = 2\omega_\text{r}$. The squeezing parameter satisfies $|M|\leq \sqrt{N(N+1)}$, with $M=0$ for a thermal reservoir and $|M|=\sqrt{N(N+1)}$ for an ideal squeezed reservoir.

The relevant detunings shown in Fig. \ref{fig:system}(a) are described as follows. $\Delta_{\text{c}}=\omega_{\text{c}}-\omega_0$ denotes the detuning of the standing-wave control field from the atomic transition frequency $\omega_0$, $\Delta_{\text{p}}=\omega_{\text{p}}-\omega_{\text{0}}$ is the detuning of the probe field relative to the control field, where  positive (negative) detuning corresponds to blue (red) detuning of the corresponding field. $\Delta=\Delta_{\text{c}}-\Delta_{\text{p}}$ denotes the overall probe detuning from the atomic resonance.
The reservoir detuning is $\delta = \omega_\text{r}-\omega_{\text{c}}$. This parameter governs the relative phase evolution between the coherent control field and the phase-sensitive reservoir correlations.
For the one-dimensional grating considered in the main text, the control-field Rabi frequency is $\Omega_{\text{c}}(x)=\Omega_{\text{c}0}\sin(\pi x / \Lambda_x)$. For the two-dimensional diffraction maps, the control field is generalized to two orthogonal standing-wave components,
\begin{equation}
	\Omega_{\text{c}}(x,y)= \Omega_{\text{c}0} \left[ \sin\!\left(\frac{\pi x}{\Lambda_x}\right) + \sin\!\left(\frac{\pi y}{\Lambda_y}\right) \right],
	\label{eq:Omega_2D}
\end{equation}
where $\Lambda_x$ and $\Lambda_y$ are the grating periods along $x$ and $y$. The parameters used in this work are listed and described in the Appendix~\ref{App:A}.

\subsection{Reservoir-modified Bloch equations}

Transforming to a frame rotating at the control-field frequency $\omega_{\text{c}}$ and employing the Born--Markov and rotating-wave approximations yields the reservoir-modified master equation for the reduced atomic density operator. The corresponding equations of motion for the expectation values of the atomic operators are the Bloch equations:
\begin{align}
	\frac{\partial}{\partial t} \langle \hat{\sigma}^{+}(t)\rangle
	&= -(\Gamma+i\Delta_{\text{c}}) \langle \hat{\sigma}^{+}(t)\rangle
	+ \gamma M e^{2i\delta t} \langle \hat{\sigma}^{-}(t)\rangle
	\nonumber\\
	&\quad + 2i\Omega_{\text{c}}(x)e^{-\eta t} \langle \hat{\sigma}^{z}(t)\rangle,
	\label{eq:bloch_plus}
	\\
	\frac{\partial}{\partial t} \langle \hat{\sigma}^{z}(t)\rangle
	&= i\Omega_{\text{c}}(x)e^{-\eta t} \left[ \langle \hat{\sigma}^{+}(t)\rangle - \langle \hat{\sigma}^{-}(t)\rangle \right]
	\nonumber\\
	&\quad - 2\Gamma \langle \hat{\sigma}^{z}(t)\rangle - \frac{\gamma}{2}.
	\label{eq:bloch_z}
\end{align}
Here $\hat{\sigma}_{\pm}$ and $\hat{\sigma}_{z}$ are pseudospin operators satisfying $[\hat{\sigma}_{+},\hat{\sigma}_{-}]=2\hat{\sigma}_{z}$ and $[\hat{\sigma}_{\pm},\hat{\sigma}_{z}] = \mp \hat{\sigma}_{\pm}$. A finite reservoir occupation $N$ enhances stimulated processes, while in the squeezed-vacuum case the correlation parameter $M$ introduces additional phase-sensitive couplings. The factor $e^{2i\delta t}$ accounts for the phase evolution associated with $\delta$, and $e^{-\eta t}$ describes a finite-duration or weakly decaying control field.

Equations~(\ref{eq:bloch_plus}) and (\ref{eq:bloch_z}) contain the following limiting cases: NV ($N=0,\;M=0$), TF ($N\neq0,\;M=0$), and SV ($N\neq0,\;M\neq0$).

\subsection{Linear susceptibility and transmission function}

The linear susceptibility of the medium is obtained from the Laplace-transformed two-time atomic correlation function,
\begin{equation}
	\chi(\Delta_{\text{p}}) = \int_{0}^{\infty} \left\langle \left[ \hat{\sigma}^{-}(t+\tau), \hat{\sigma}^{+}(\tau) \right] \right\rangle e^{-i\Delta_{\text{p}}\tau} \,d\tau .
	\label{eq:chi_def}
\end{equation}
where $\tau$ denotes the delay time between the two atomic operators.
Introducing the Laplace variable $c=i\Delta_{\text{p}}$, the susceptibility can be written compactly as  
\begin{equation}
	\chi(c) = -2u_{+}^{*}(c)\hat{\sigma}^{z} + u_{z}^{*}(c)\hat{\sigma}^{+},
\label{eq:chi_compact}
\end{equation}
 where the auxiliary response functions $u_{+}^{*}(c)$ and $u_z^{*}(c)$ contain the complete dependence on the reservoir parameters ($N$, $M$, $\delta$, $r$, $\eta$) and are given in Appendix~\ref{App:B}.

Although Eqs.~(\ref{eq:bloch_plus}) and (\ref{eq:bloch_z}) contain explicit time-dependent factors, the diffraction calculations use the weak-driving Laplace-domain response. The temporal dependence is absorbed into the complex coefficients through $\kappa=\eta+i\delta$ and $\kappa^{*}=\eta-i\delta$, which appear in the expressions for $u_{+}^{*}(c)$ and $u_z^{*}(c)$. Thus the calculated susceptibility is an effective stationary quantity $\chi(\Delta_{\text{p}})$, describing the steady-state or quasi-stationary reservoir-modified EIG response. We decompose the susceptibility as $\chi(\Delta_{\text{p}})=\chi'(\Delta_{\text{p}})+i\chi''(\Delta_{\text{p}})$, where $\chi'=\operatorname{Re}\chi$ is the dispersive part and $\chi''=\operatorname{Im}\chi$ is the absorptive (gain) part. 

Under the slowly varying envelope approximation and the paraxial approximation, the propagation equation for the weak probe field is given by \cite{caves1985formalism1}
\begin{equation}
	\frac{\partial E_{\text{p}}}{\partial z} + \frac{1}{c} \frac{\partial E_{\text{p}}}{\partial t} = i \frac{\pi}{\lambda_{\text{p}}} \chi E_{\text{p}},
\end{equation}
where $\lambda_{\text{p}}$ is the probe wavelength. Considering the steady-state response and introducing the dimensionless coordinate $z' = z/z_0$ with $z_0 = \lambda_{\text{p}}/(2\pi |\boldsymbol{\mu}_{eg}|^2 \mathcal{N}_0)$ (where $\mathcal{N}_0$ is the atomic density), the propagation equation reduces to
\begin{equation}
	\frac{\partial E_{\text{p}}}{\partial z'} = i \chi E_{\text{p}}.
	\label{eq:propagation_dimensionless}
\end{equation}
For notational simplicity, we drop the prime and use $z$ as the dimensionless propagation distance.
Integrating Eq.~(\ref{eq:propagation_dimensionless}) across a medium of thickness $L/z_0$ (in dimensionless units) yields the transmission function
\begin{equation}
	T(x) = \exp\left[ i \chi(x) L/z_0 \right] = \exp\left[ -\chi''(x)L/z_0 + i \chi'(x)L/z_0 \right],
	\label{eq:T_1D}
\end{equation}
where $\chi = \chi' + i \chi''$ is the linear susceptibility, with $\chi' = \operatorname{Re}\chi$ the dispersive component and $\chi'' = \operatorname{Im}\chi$ the absorptive (gain) component.
The factor $e^{-\chi''(x)L/z_0}$ describes spatial gain (if $\chi'' < 0$) or attenuation (if $\chi'' > 0$), while $e^{i\chi'(x)L/z_0}$ describes spatial phase modulation. Their combined action forms a gain-phase electromagnetically induced grating. The phase shift accumulated after propagation is $\Phi(x) = \chi'(x)L/z_0$, and the amplitude modulation depth is governed by $\chi''(x)L/z_0$.
For the two-dimensional geometry, the transmission function generalizes to
\begin{equation}
	T(x,y) = \exp\left[ -\chi''(x,y)L/z_0 + i \chi'(x,y)L/z_0 \right].
	\label{eq:T_2D}
\end{equation}

\subsection{Fraunhofer diffraction}

For a one-dimensional transmission grating, the single-period Fraunhofer diffraction reads
\begin{equation}
	F(\theta) = \int_0^1 T(x) \exp\!\left[-i2\pi x\sin\theta\right] dx.
	\label{eq:F_1D}
\end{equation}
The normalized intensity can be accordingly introduced as $I_{\text{p}}(\theta)=|F(\theta)|^2$.
For the two-dimensional grating, the amplitude is obtained from the two-dimensional Fourier transform
\begin{equation}
	\begin{split}
		F(\theta_x,\theta_y) &= \int_0^1\int_0^1 T(x,y) \\
		&\quad\times \exp\left[ -i2\pi \left( x\sin\theta_x + y\sin\theta_y \right) \right] dx\,dy ,
	\end{split}
	\label{eq:F_2D}
\end{equation}
For a finite grating containing $\mathcal{N}_x$ illuminated periods along $x$ and $\mathcal{N}_y$ along $y$, the normalized intensity is
\begin{align}
	I_{\text{p}}(\theta_x,\theta_y)
	&= \left|F(\theta_x,\theta_y)\right|^2
	\frac{ \sin^2 \left( \mathcal{N}_x\pi \frac{\Lambda_x}{\lambda_{\text{p}}} \sin\theta_x \right) }{ \mathcal{N}_x^2 \sin^2 \left( \pi \frac{\Lambda_x}{\lambda_{\text{p}}} \sin\theta_x \right) }
	\nonumber\\
	&\quad\times \frac{ \sin^2 \left( \mathcal{N}_y\pi \frac{\Lambda_y}{\lambda_{\text{p}}} \sin\theta_y \right) }{ \mathcal{N}_y^2 \sin^2 \left( \pi \frac{\Lambda_y}{\lambda_{\text{p}}} \sin\theta_y \right) },
	\label{eq:I_2D}
\end{align}
where $\lambda_{\text{p}}$ is the probe wavelength. The principal diffraction maxima occur at
\begin{equation}
	\sin\theta_x=\frac{m\lambda_{\text{p}}}{\Lambda_x},\quad \sin\theta_y=\frac{n\lambda_{\text{p}}}{\Lambda_y},\quad m,n=0,\pm1,\pm2,\ldots \nonumber.
	\label{eq:orders_2D}
\end{equation}
Equations~(\ref{eq:T_1D})--(\ref{eq:I_2D}) connect the reservoir-modified susceptibility to the observable diffraction spectra. Parameters that increase the spatial modulation depth of $T(x)$ or $T(x,y)$ strengthen the relevant Fourier components and enhance the diffracted orders, whereas parameters that wash out the spatial contrast suppress the diffraction intensity.

%%%%%%%%%%%%%%%%%%%%%%%%%%%%%%%%%
\section{Results and Discussion}\label{sec:results}
%%%%%%%%%%%%%%%%%%%%%%%%%%%%%
We now investigate how engineered reservoirs modify the transmission and diffraction properties of the probe field in an EIG. In this section, all frequency-like parameters are normalized to the spontaneous emission rate \(\gamma\). Unless stated otherwise, the fixed parameters in this work are \(\mathcal{R}/\gamma = 0.1\), \(\Delta_{\mathrm c}/\gamma = 0.2\), \(\Omega_{\mathrm c0}/\gamma = 0.1\), \(L/z_0 = 0.1\), \(\phi = 0\), $\mathcal{N}_x = \mathcal{N}_y = 5$, and $\Lambda_x/\lambda_\text{p} = \Lambda_y/\lambda_\text{p} = 4$. The normalized propagation length is fixed at $L/z_0 = 0.1$, which is sufficient to produce pronounced transmission modulation and well-resolved diffraction patterns in the present parameter regime. Larger lengths primarily increase diffraction efficiency without qualitatively altering the reservoir-induced effects. These values place the system in the weak-driving regime, where reservoir-induced modifications of atomic coherence can be clearly identified without entering strong saturation.
Our analysis proceeds from near-field transmission profiles to the corresponding far-field diffraction patterns. Particular attention is paid to the influence of the reservoir photon number \(N\), the control-field decay parameter \(\eta/\gamma\), and the squeezed-reservoir detuning \(\delta/\gamma\), which all these parameters are experimentally accessible for tailoring the optical response of the system.
% ----------------------------------------------
\begin{figure}[t]
	\centering
\includegraphics[scale=0.4,trim=0 0 0 0, clip]{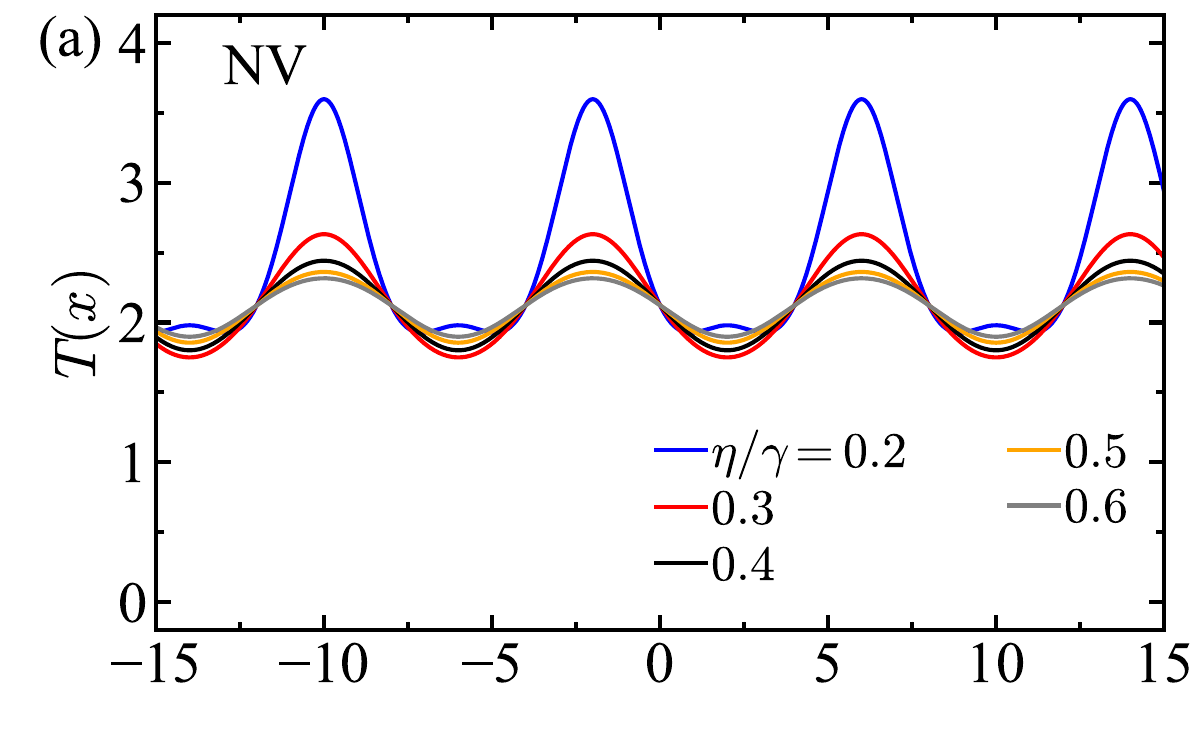}
\includegraphics[scale=0.4,trim=0 0 0 0, clip]{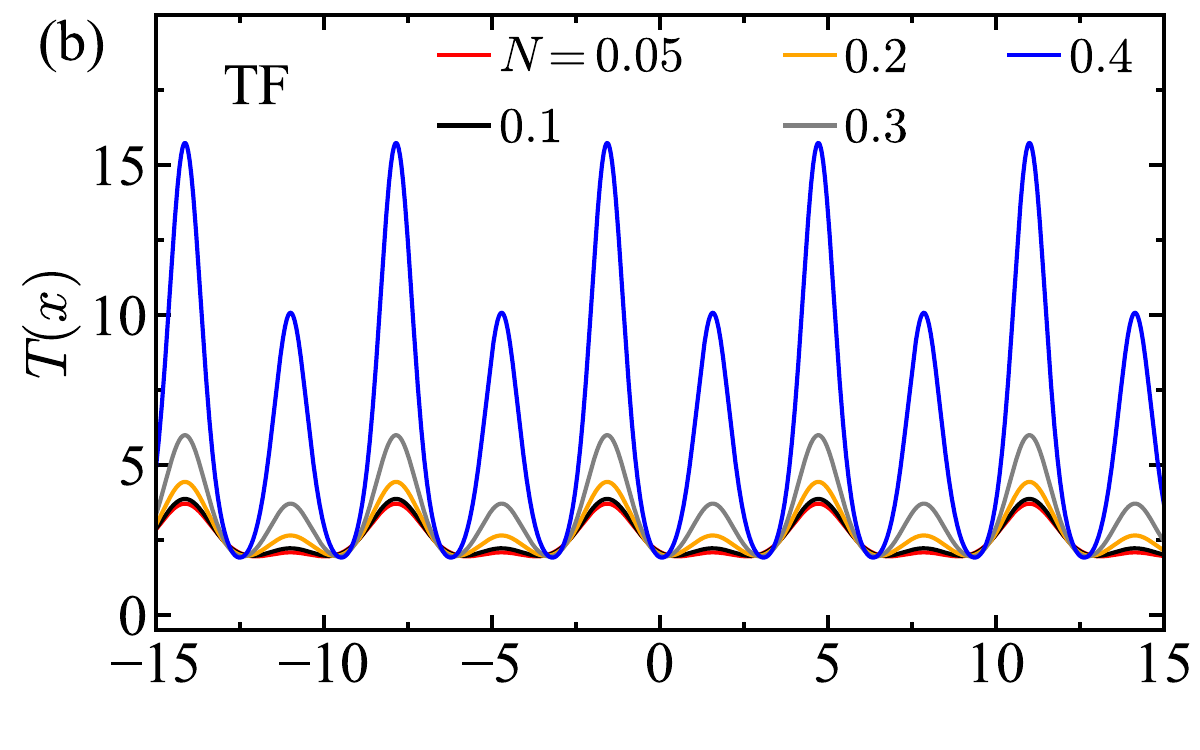}
\includegraphics[scale=0.4,trim=0 0 0 0, clip]{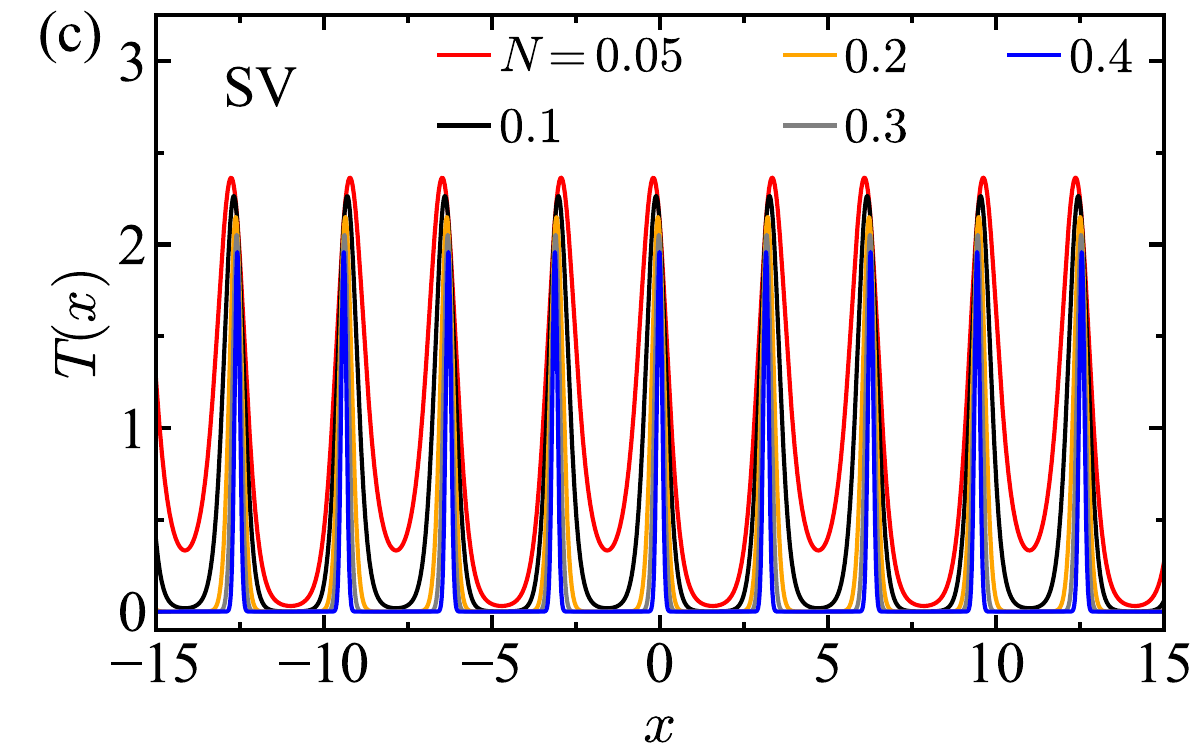}
	\caption{
		Transmission $T(x)$ of the probe field in $x$-direction as a function of the spatial coordinate $x$ for
		(a) the NV reservoir ($N=0$) for several decay parameters $\eta/\gamma = 0.2,\,0.3,\,0.4,\,0.5, 0.6$.
		(b) the TF reservoir ($M=0$) for $N = 0.05,\,0.1,\,0.2,\,0.3,\,0.4$ and fixed $\eta/\gamma = 0.2$; and
		(c) the SV reservoir for $N = 0.05,\,0.1,\,0.2,\,0.3,\,0.4$ and $\eta/\gamma = 0.2$.
		Fixed parameters are: $\mathcal{R}/\gamma=0.1$, $\Delta_\text{c}/\gamma=0.2$, $\Omega_\text{c0}/\gamma=0.1$, $L/z_0=0.1$, $\Delta_\text{p}/\gamma=\phi=0$, $\mathcal{N}_x=5$, and $\Lambda_x/\lambda_\text{p}=4$.
	} \label{fig:2}
\end{figure}

Figure~\ref{fig:2} presents the spatial transmission function \(T(x)\) for three reservoir types. In the NV case [Fig.~\ref{fig:2}(a)], the transmission exhibits a periodic modulation inherited from the standing-wave control field. Increasing the control-field decay parameter \(\eta/\gamma\)  reduces the transmission maxima while preserving the spatial period, indicating that a larger \(\eta/\gamma\) weakens the effective grating contrast experienced by the probe field.
The behavior changes significantly in the presence of a thermal reservoir. As shown in Fig.~\ref{fig:2}(b), increasing the thermal photon number \(N\) leads to a substantial enhancement of the transmission peaks. Simultaneously, the transmission profile develops sharper spatial features, reflecting a stronger modulation of the probe susceptibility. Thus, the reservoir acts as an effective amplifier of the EIG response.
The most sharp modification occurs for the squeezed-vacuum reservoir shown in Fig.~\ref{fig:2}(c). Even for moderate values of \(N\), the transmission evolves into a sequence of narrow, high-contrast peaks separated by regions of strongly reduced transmission. This behavior originates from the phase-sensitive reservoir correlations encoded in the squeezing parameter \(M = \sqrt{N(N+1)}\) (with $\phi=0$). These correlations generate a substantially stronger spatial modulation than a purely thermal reservoir. Consequently, the squeezed reservoir provides an efficient mechanism for engineering highly localized transmission channels within the grating structure.

\begin{figure}
	\centering
\includegraphics[scale=0.35,trim=0 20 0 0, clip]{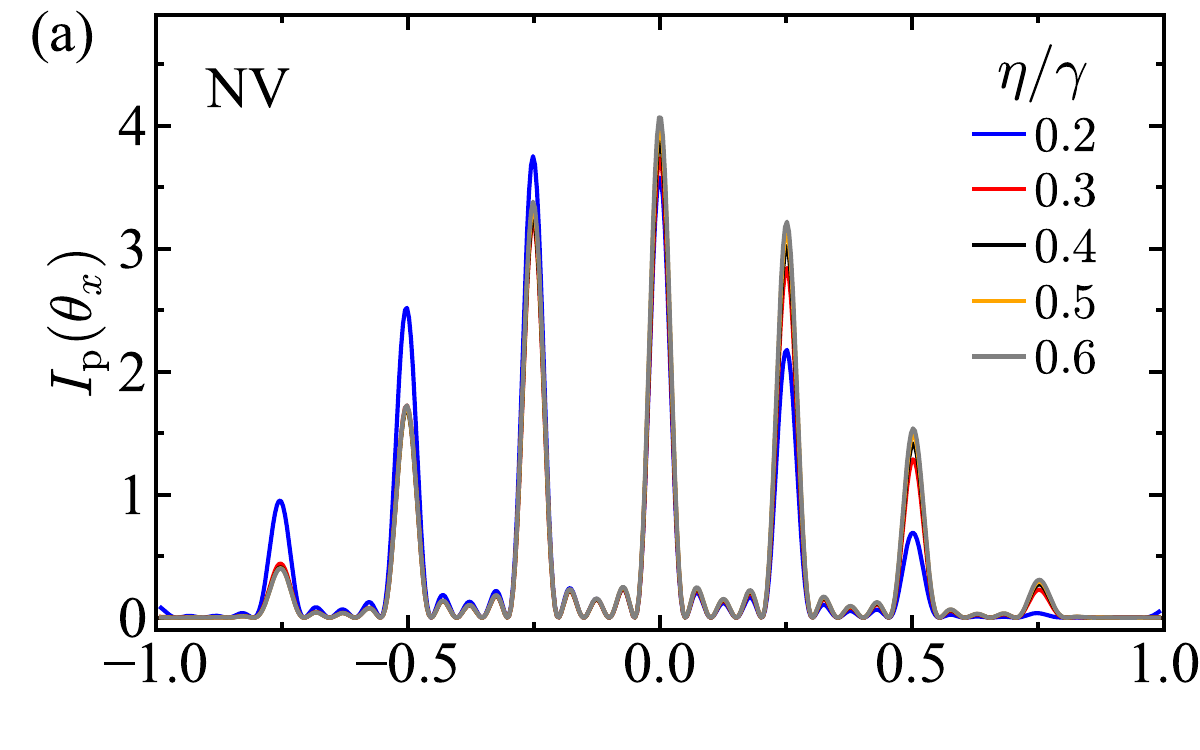}
\includegraphics[scale=0.35,trim=0 20 0 0, clip]{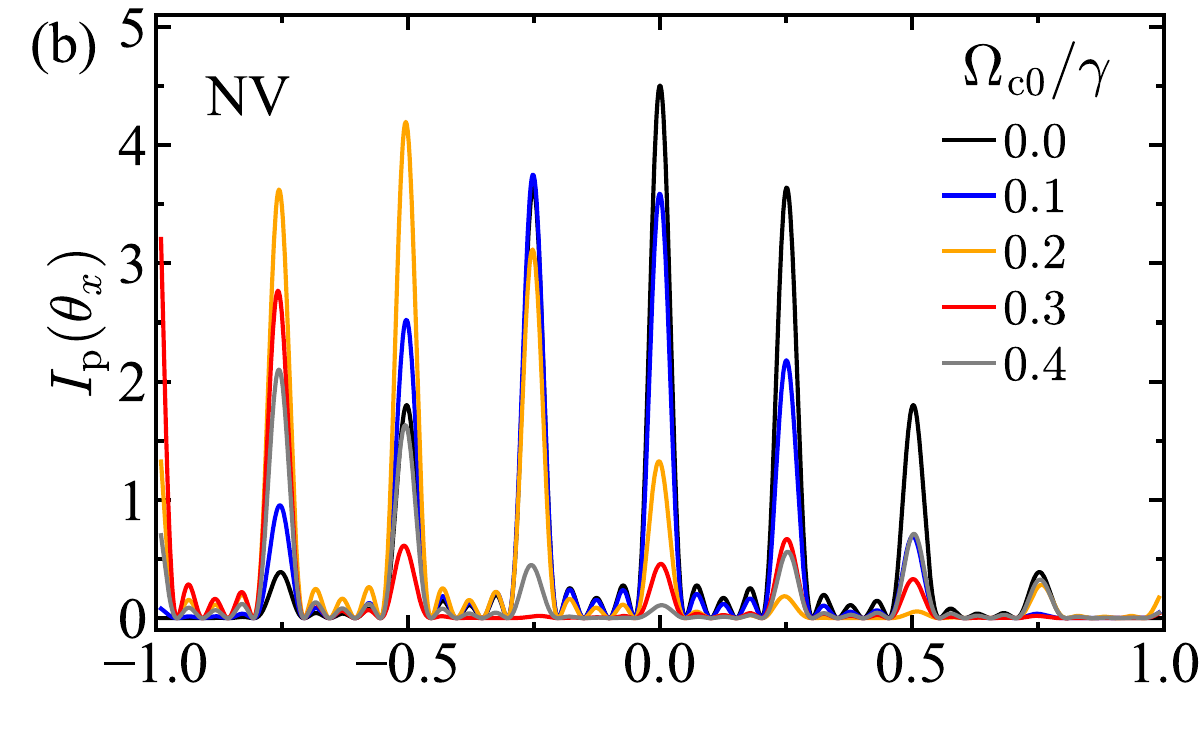}
\includegraphics[scale=0.35,trim=0 0 0 0, clip]{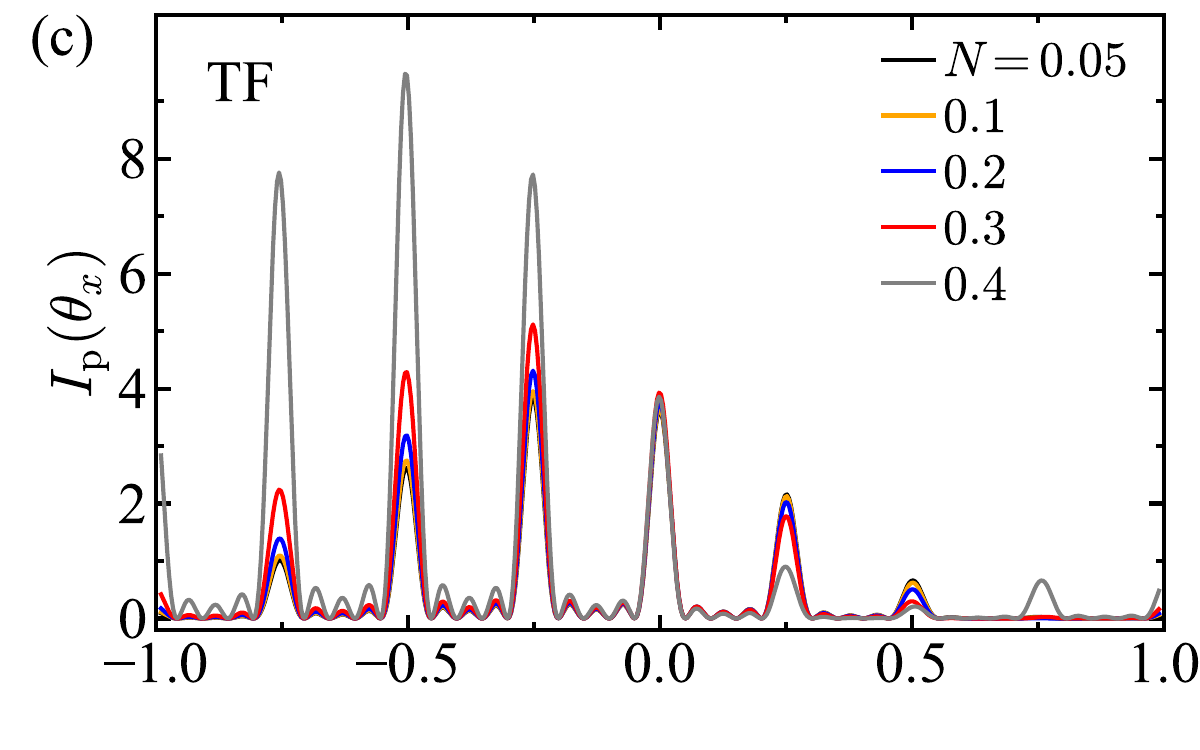}
\includegraphics[scale=0.35,trim=0 0 0 0, clip]{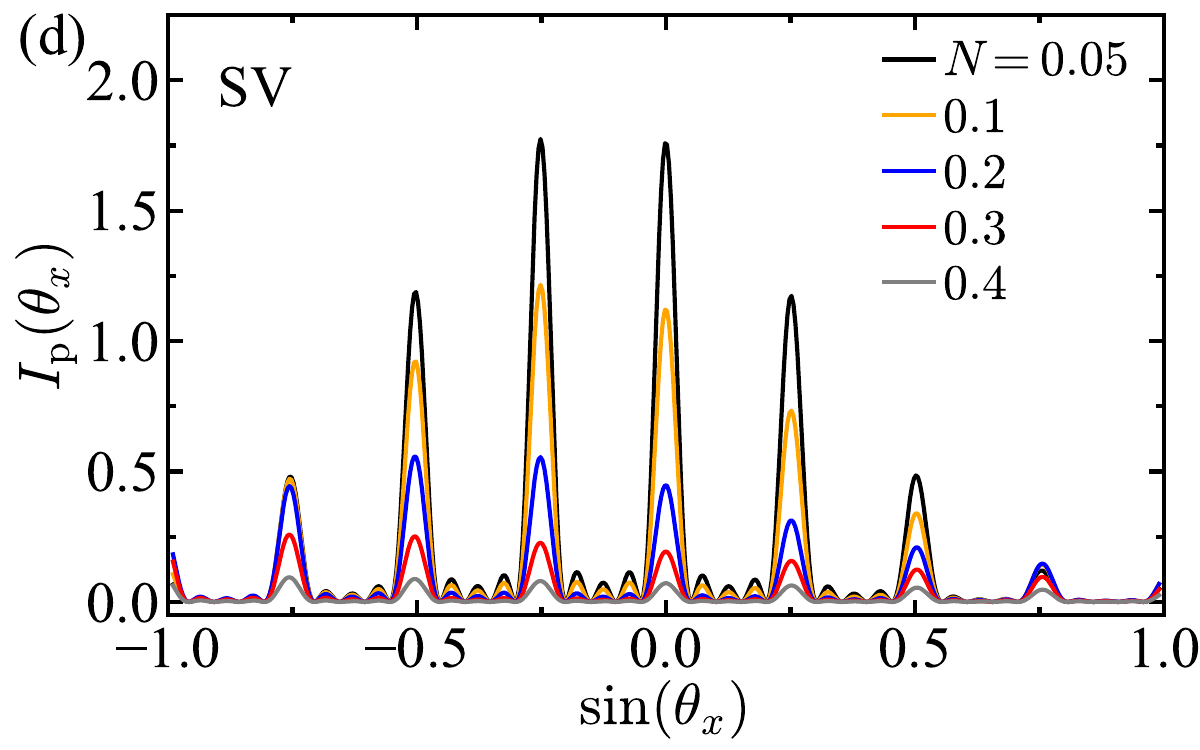}
\caption{
Normalized diffraction intensity $I_\text{p}(\theta_x)$ as a function of $\sin(\theta_x)$.
(a) the NV reservoir for several decay parameters $\eta/\gamma = 0.2,\,0.3,\,0.4,\,0.5, 0.6$.
(b) the NV reservoir for fixed $\eta/\gamma = 0.2$ and  several values of the Rabi frequency $\Omega_\text{c0}/\gamma = 0.0,\,0.1,\,0.2,\,0.3, 0.4$.
(c), (d) The TF and SV reservoirs for $N = 0.05,\,0.1,\,0.2,\,0.3,\,0.4$, assuming $\eta/\gamma = 0.2$ and $\Omega_\text{c0}/\gamma =0.1$.
Fixed parameters are:$\mathcal{R}/\gamma=0.1$, $\Delta_\text{c}/\gamma=0.2$, $\Omega_\text{c0}/\gamma=0.1$, $L/z_0=0.1$, $\Delta_\text{p}/\gamma=\phi=0$, $\mathcal{N}_x=5$, and $\Lambda_x/\lambda_\text{p}=4$.
}
\label{fig:3}
\end{figure}

The corresponding one-dimensional diffraction intensity $I_\text{p}(\theta_x)$ is shown in Fig.~\ref{fig:3} under different circumstances. In all cases, the diffraction pattern consists of discrete angular orders arising from the Fourier components of the transmission grating.
For the NV reservoir [Fig.~\ref{fig:3}(a)], increasing \(\eta/\gamma\) progressively suppresses the diffraction orders, consistent with the reduction in transmission modulation observed in Fig.~\ref{fig:2}(a). Figure~\ref{fig:3}(b) shows that increasing the control-field amplitude \(\Omega_{\mathrm c0}\) reshapes the diffraction peaks particularly for stronger values of $\delta/\gamma$ (see orange curve) it enhances higher-order diffraction channels, reflecting the larger modulation depth of the induced grating.
As depicted in Fig.~\ref{fig:3}(c), the TF reservoir substantially increases the intensity of the diffracted orders as \(N\) increases. The enhancement is particularly pronounced for the principal diffraction maxima, demonstrating that thermal photons can strengthen the spatial Fourier components of the susceptibility.
The SV reservoir [Fig.~\ref{fig:3}(d)] exhibits qualitatively different behavior. Although the diffraction intensity is enhanced, the amplification is not distributed uniformly among all diffraction orders. Instead, selected angular channels become dominant while others remain comparatively weak. This selective enhancement reflects the phase-sensitive nature of the squeezed reservoir and provides an additional degree of control over the angular distribution of the probe field.

\begin{figure}[t]
	\centering
\includegraphics[scale=0.4,trim=0 0 0 0, clip]{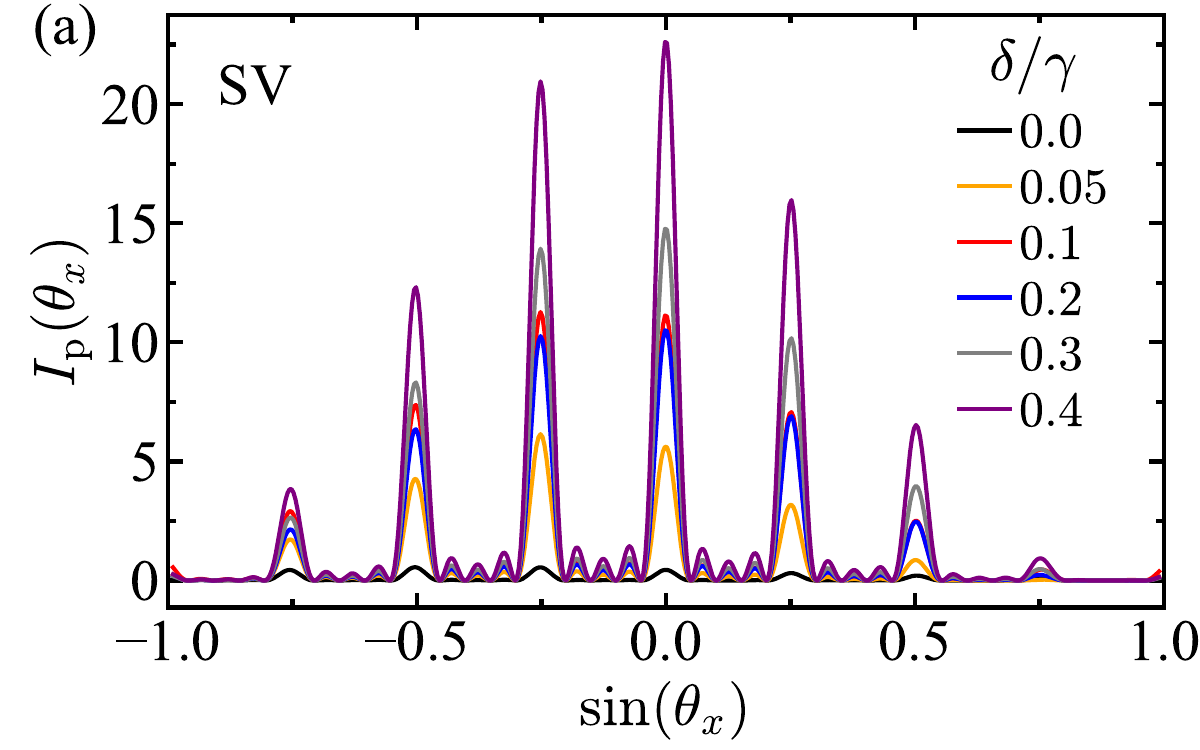}
\includegraphics[scale=0.4,trim=0 0 0 0, clip]{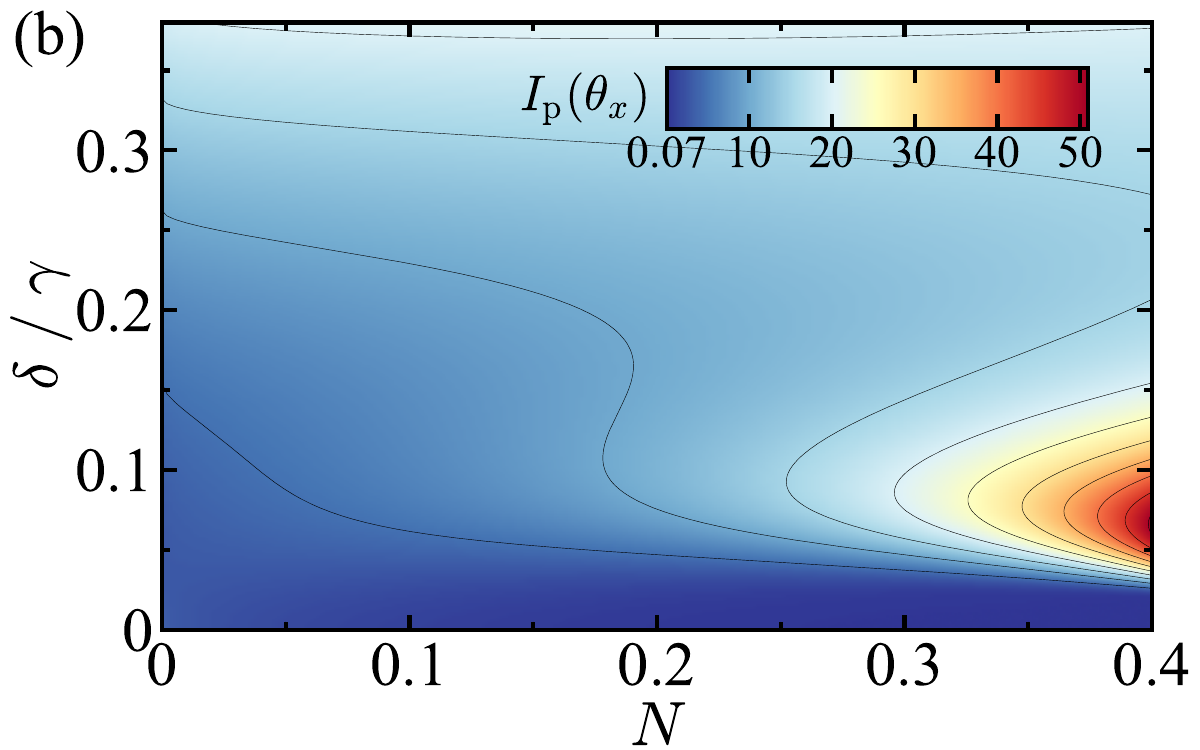}

\caption{
Diffraction intensity for the SV reservoir with nonzero squeezed-vacuum detuning.
(a)  $I_\text{p}(\theta_x)$ versus $\sin(\theta_x)$ for several values of $\delta/\gamma$ at fixed $N=0.2$.
(b) Contour plot of $I_\text{p}(\theta_x)$ in the $\delta/\gamma$-$N$ plane evaluated at 
$\sin(\theta)=0$, showing the parameter region of maximum amplification.
Other fixed parameters are chosen as Fig. \ref{fig:3}.
}
\label{fig:4}
\end{figure}

To further explore the role of phase-sensitive reservoir engineering, Fig.~\ref{fig:4} investigates the effect of a nonzero squeezed-reservoir detuning \(\delta/\gamma\).
Figure~\ref{fig:4}(a) shows the diffraction intensity \(I_p(\theta_x)\) for several values of \(\delta/\gamma\) at fixed \(N=0.2\). Increasing the detuning strongly enhances the dominant diffraction orders while preserving their angular locations. The enhancement is highly nonuniform, indicating that the detuning modifies the relative weights of the Fourier components rather than simply scaling the overall diffraction intensity.
In Fig.~\ref{fig:4}(b), we display the contour plot of the  diffraction intensity  in the \((N,\delta/\gamma)\) plane to demonstrate the parameter dependence of the intensity with more details. The strongest amplification occurs in the region of large reservoir photon number and small positive detuning. In particular, the maximum diffraction efficiency is concentrated near \(N\approx0.35-0.4\) and \(\delta/\gamma\approx0.05-0.1\). This result demonstrates that the combined action of squeezing and detuning provides an effective mechanism for enhancing the diffraction response.
As a consequence, the squeezed-reservoir detuning modifies the phase-sensitive interference processes that contribute to the atomic coherence. Also, we observe that the susceptibility acquires a stronger spatial modulation, leading to enhanced diffraction efficiency and greater control over the angular distribution of the probe field.

 \begin{figure*}[t]
	\centering
	\includegraphics[scale=0.29,trim=0 0 0 0, clip]{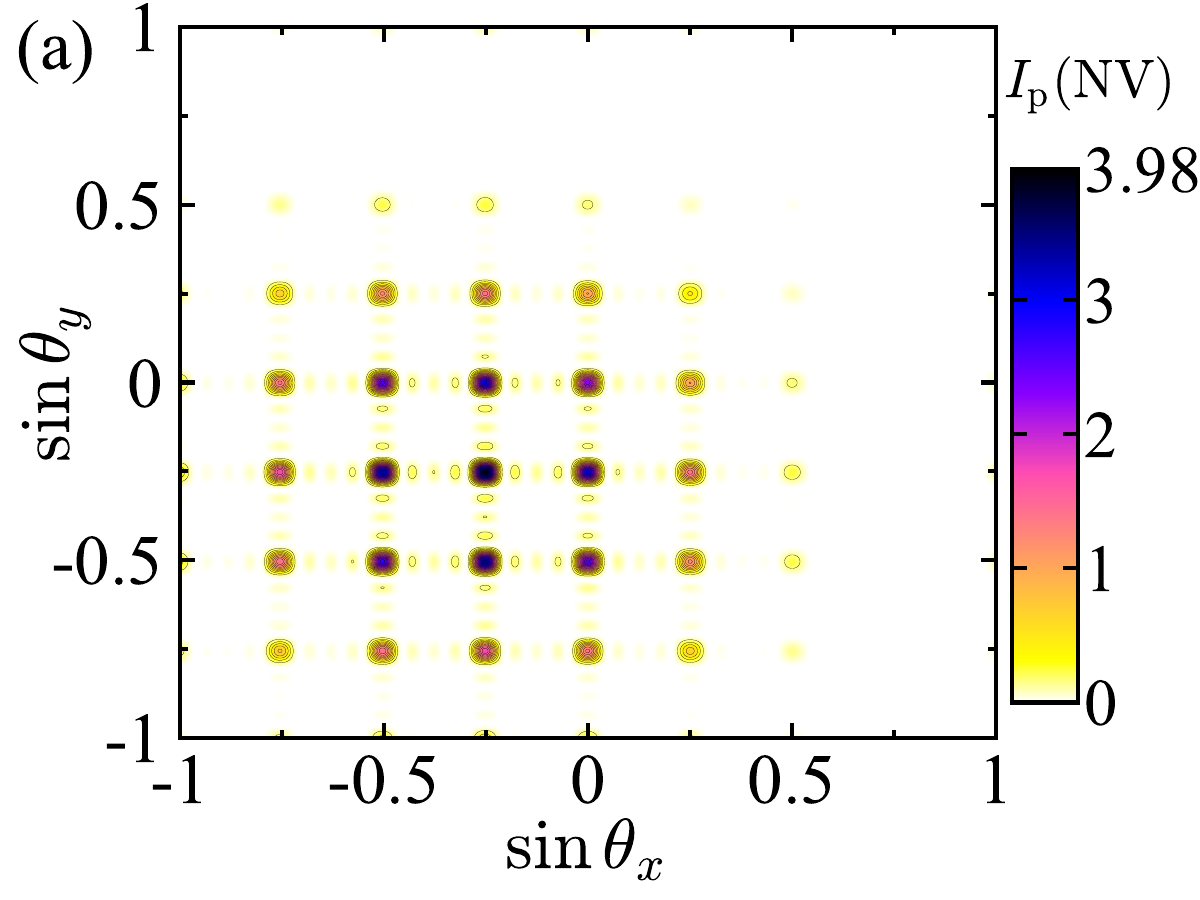}
    \includegraphics[scale=0.29,trim=0 0 0 0, clip]{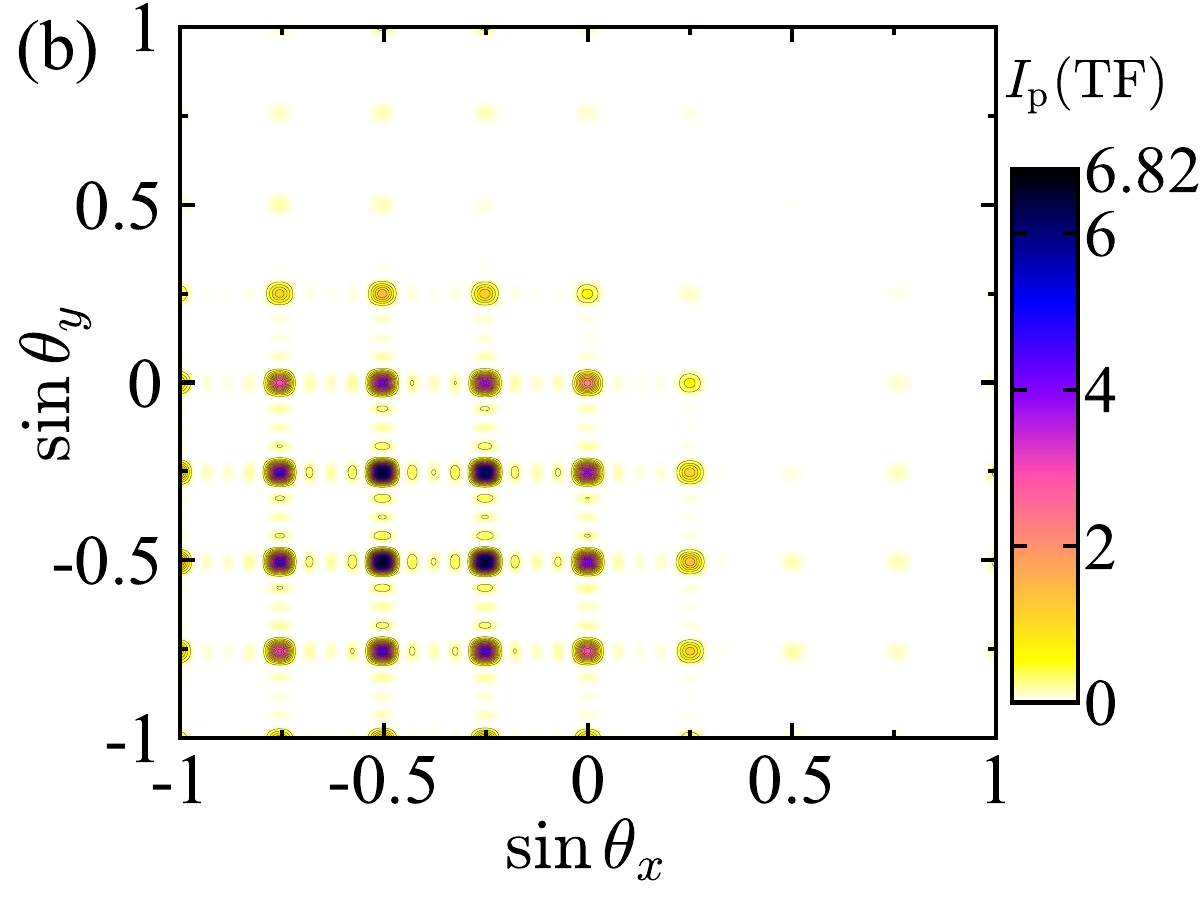}
    \includegraphics[scale=0.29,trim=0 0 0 0, clip]{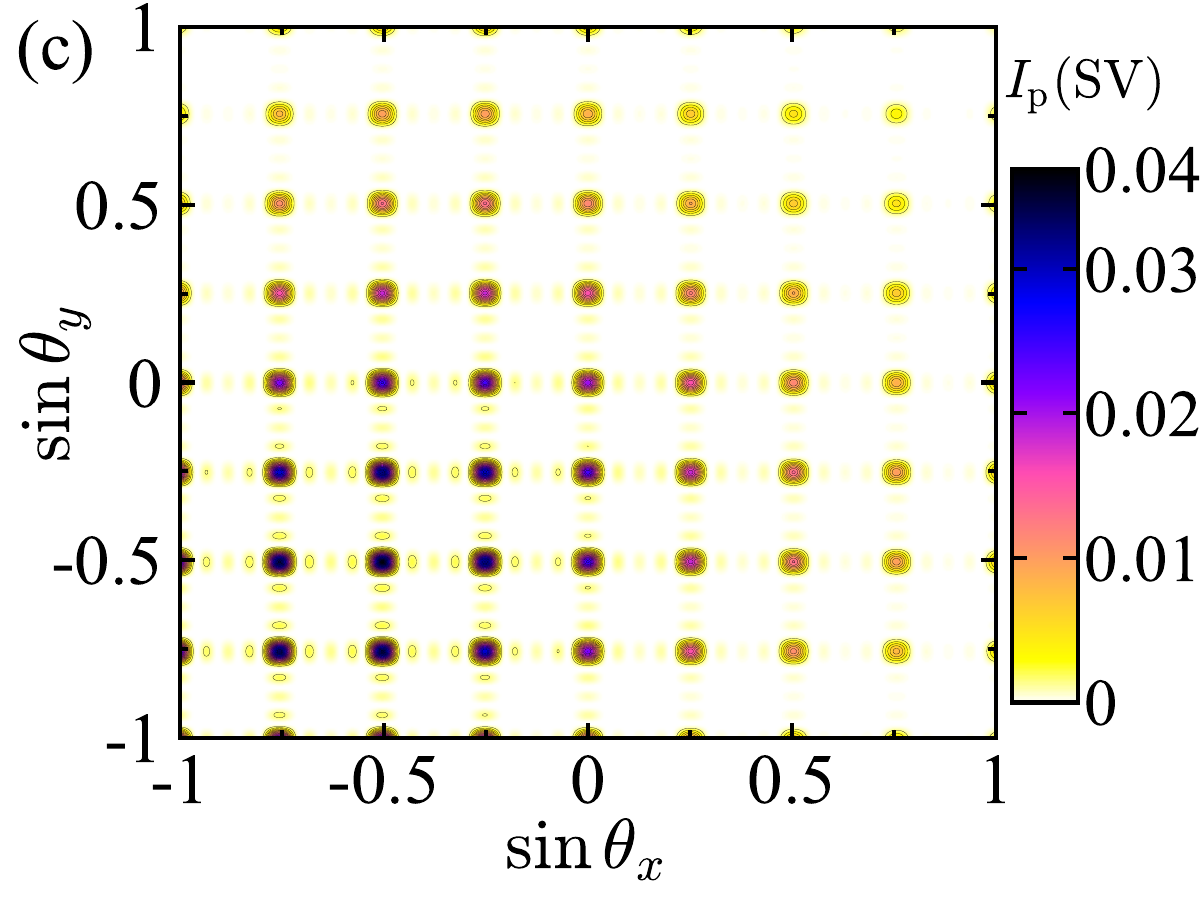}
    \includegraphics[scale=0.29,trim=0 0 0 0, clip]{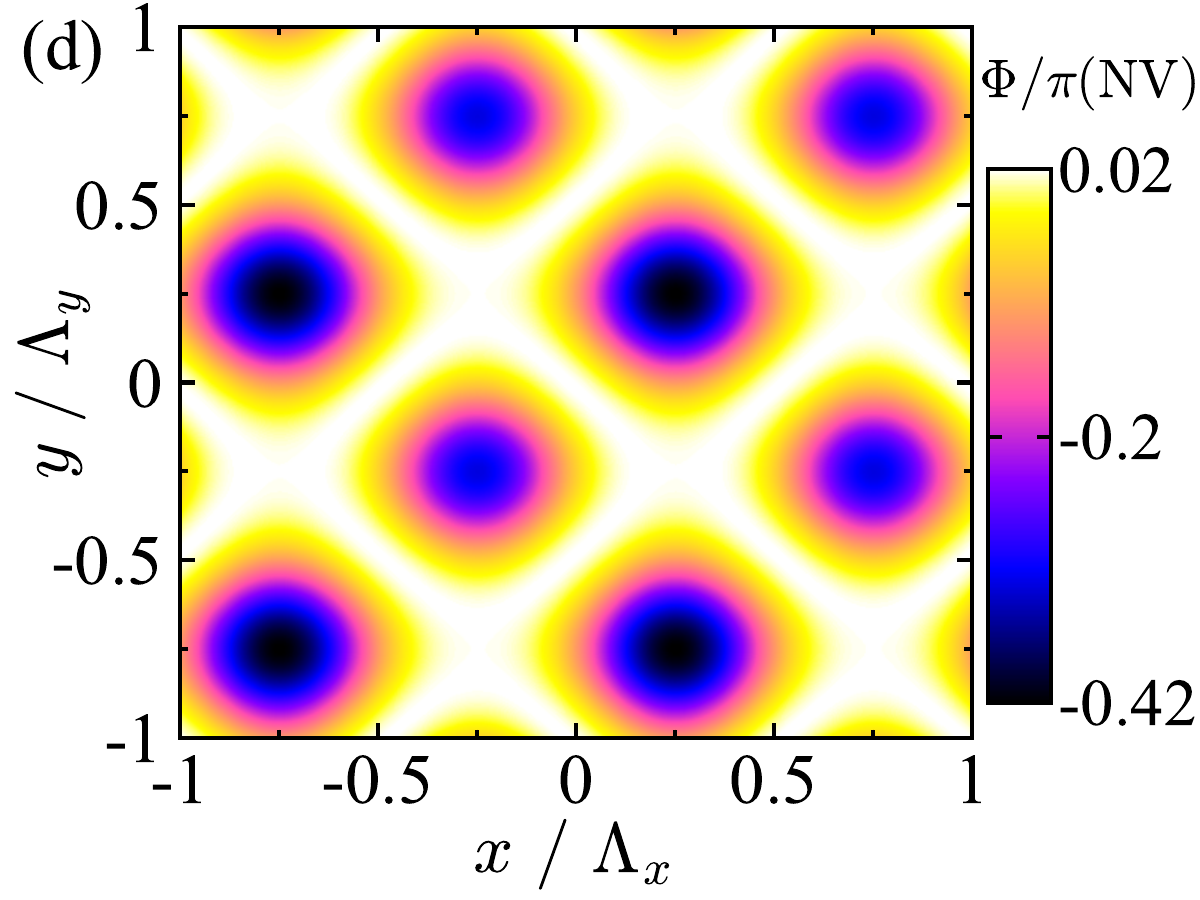}
     \includegraphics[scale=0.29,trim=0 0 0 0, clip]{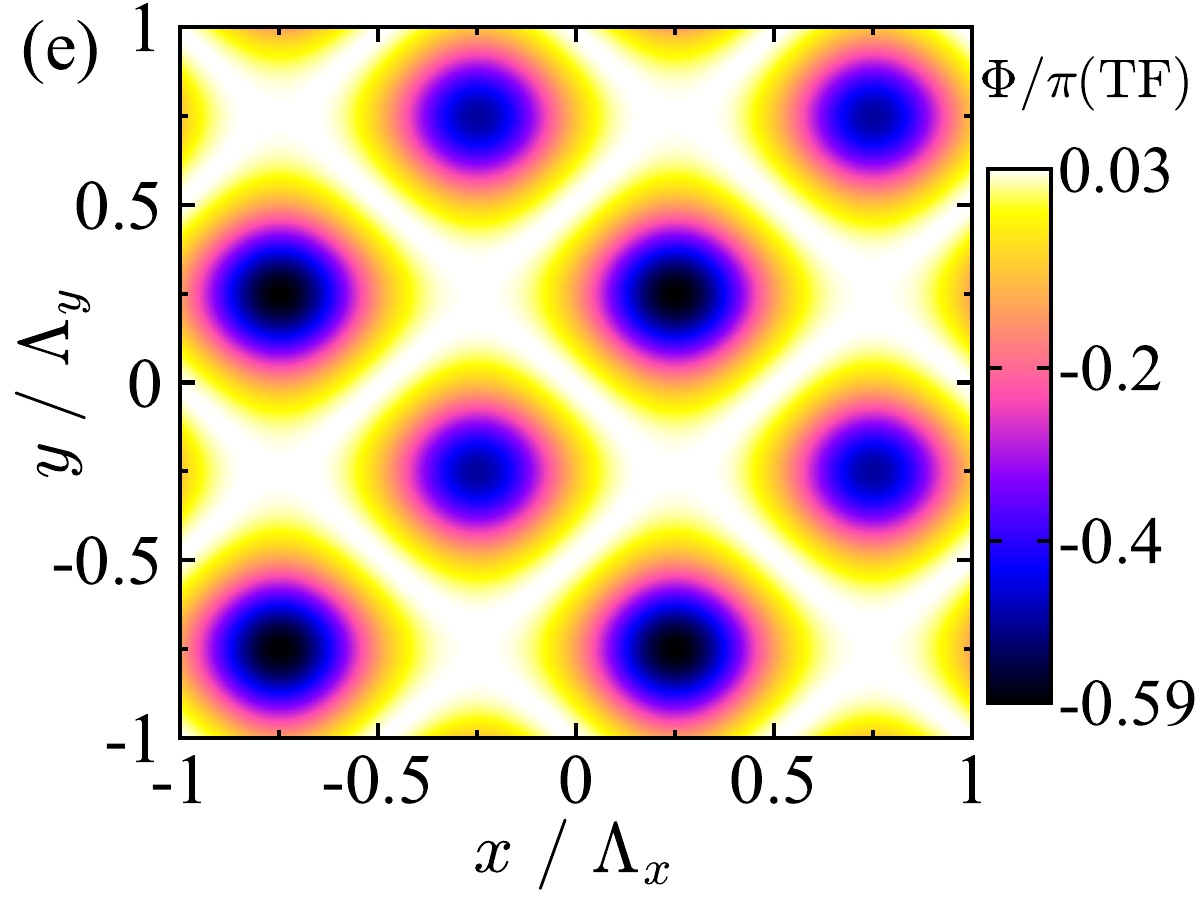}
    \includegraphics[scale=0.29,trim=0 0 0 0, clip]{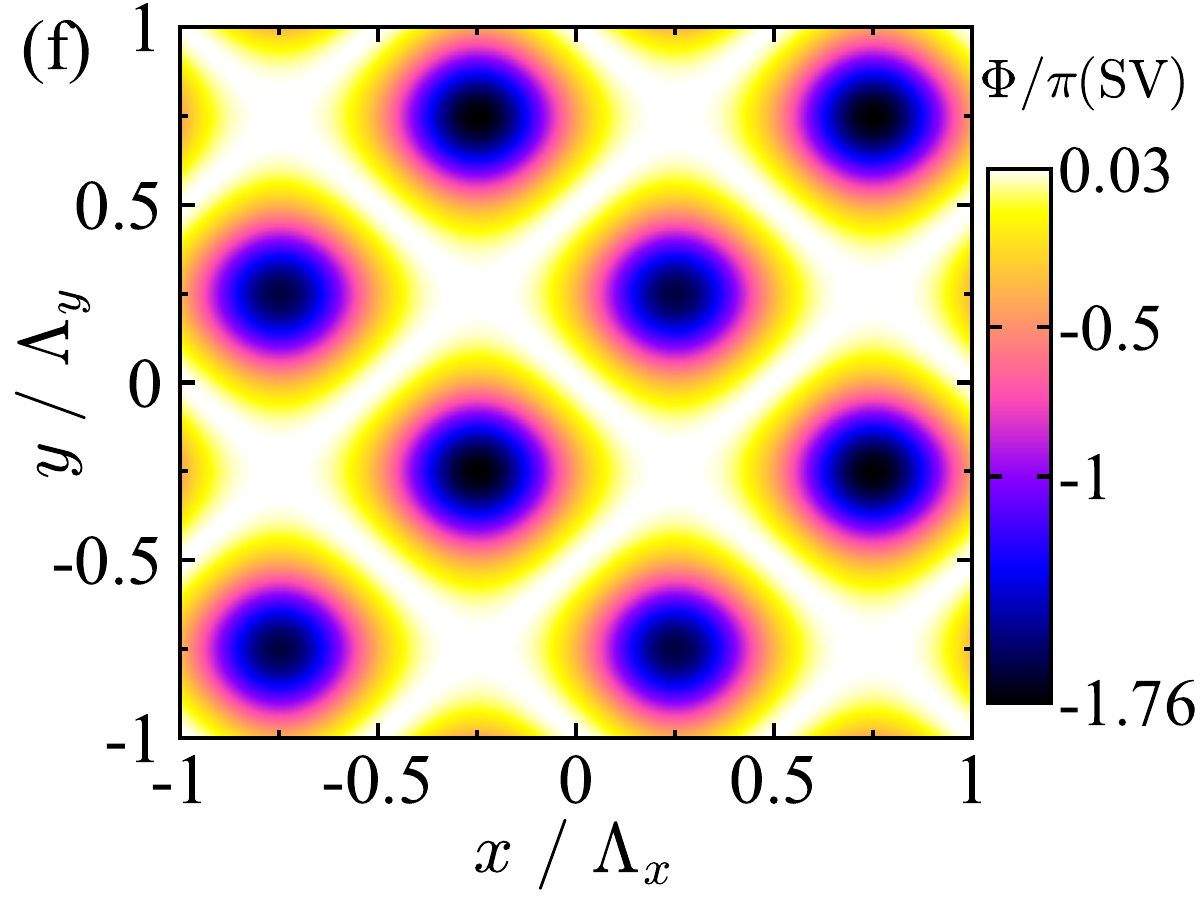}
\caption{(a), (b), (c) 
Two-dimensional diffraction intensity $I_\text{p}(\theta_x,\theta_y)$ as a function of $\sin(\theta_x)$ and $\sin(\theta_y)$ for NV, TF ($N=0.2$), SV ($N=0.2$).
 (d), (e), (f) the corresponding contour plot of phase $\Phi/\pi$ for NV, TF, SV reservoir.
Other parameters are assumed as follows: $\mathcal{R}/\gamma=0.1$, $\Delta_\text{c}/\gamma=0.2$, $\Omega_\text{c0}/\gamma=0.1$, $L/z_0=0.1$,  $\Delta_\text{p}/\gamma=\phi=0$, $\mathcal{N}_x = \mathcal{N}_y = 5$ and $\Lambda_x/\lambda_\text{p} = \Lambda_y/\lambda_\text{p} = 4$.
}
\label{fig:5}
\end{figure*}

The two-dimensional diffraction patterns are presented in Fig.~\ref{fig:5} for zero squeezed-reservoir detuning (\(\delta/\gamma=0\)). Panels \ref{fig:5}(a)-(c) show the diffraction intensity \(I_p(\theta_x,\theta_y)\), while panels \ref{fig:5}(d)-(f) display the corresponding phase distributions \(\Phi(x,y)\).
For all reservoirs, the diffraction pattern consists of a two-dimensional lattice of discrete diffraction orders, reflecting the periodic modulation imposed by the standing-wave control fields along both transverse directions (see also Fig. \ref{fig:system}(b)). The NV reservoir exhibits a relatively uniform distribution of diffraction intensity among the accessible diffraction channels [Fig. \ref{fig:5}(a)]. The TF reservoir increases the overall diffraction strength while preserving the general structure of the diffraction lattice [Fig. \ref{fig:5}(b)].
On the other hand, the SV reservoir exhibits a distinctly different behavior [Fig. \ref{fig:5}(c)]. Namely, the diffraction intensity becomes more strongly concentrated into selected channels, indicating that the squeezed reservoir redistributes optical power among the available diffraction orders.  Compared with the NV and TF cases [Figs. \ref{fig:5}(d)-(e)], as shown in Fig.~\ref{fig:5}(f), the SV reservoir produces a much larger phase excursion across the transverse plane, revealing the strong influence of phase-sensitive quantum correlations on the optical response. The redistribution of the diffraction intensity for the SV case is accompanied by a substantially larger phase modulation of the transmitted field, as evidenced by the phase maps illustrated in Fig.~\ref{fig:5}(f).

\begin{figure*}[t]
	\centering
	\includegraphics[scale=0.29,trim=0 0 0 0, clip]{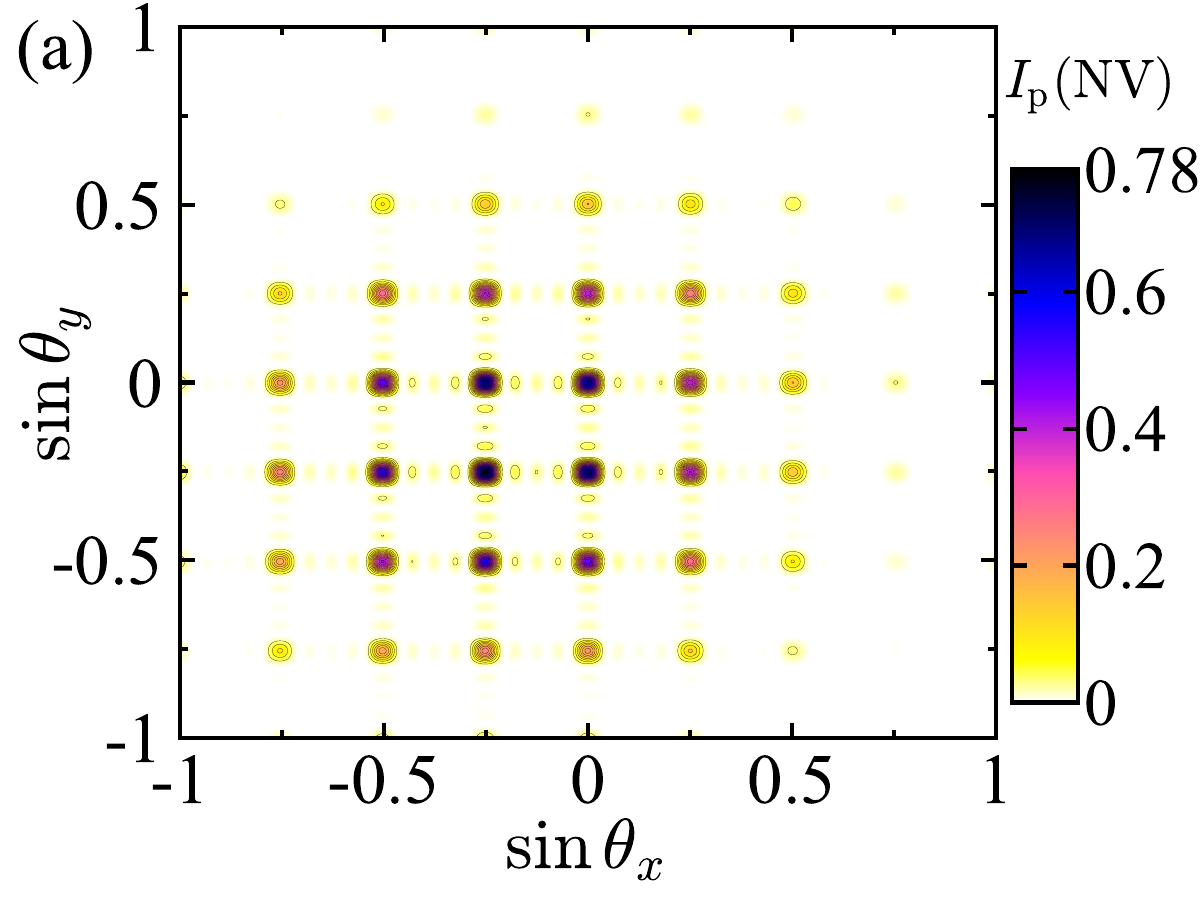}
    \includegraphics[scale=0.29,trim=0 0 0 0, clip]{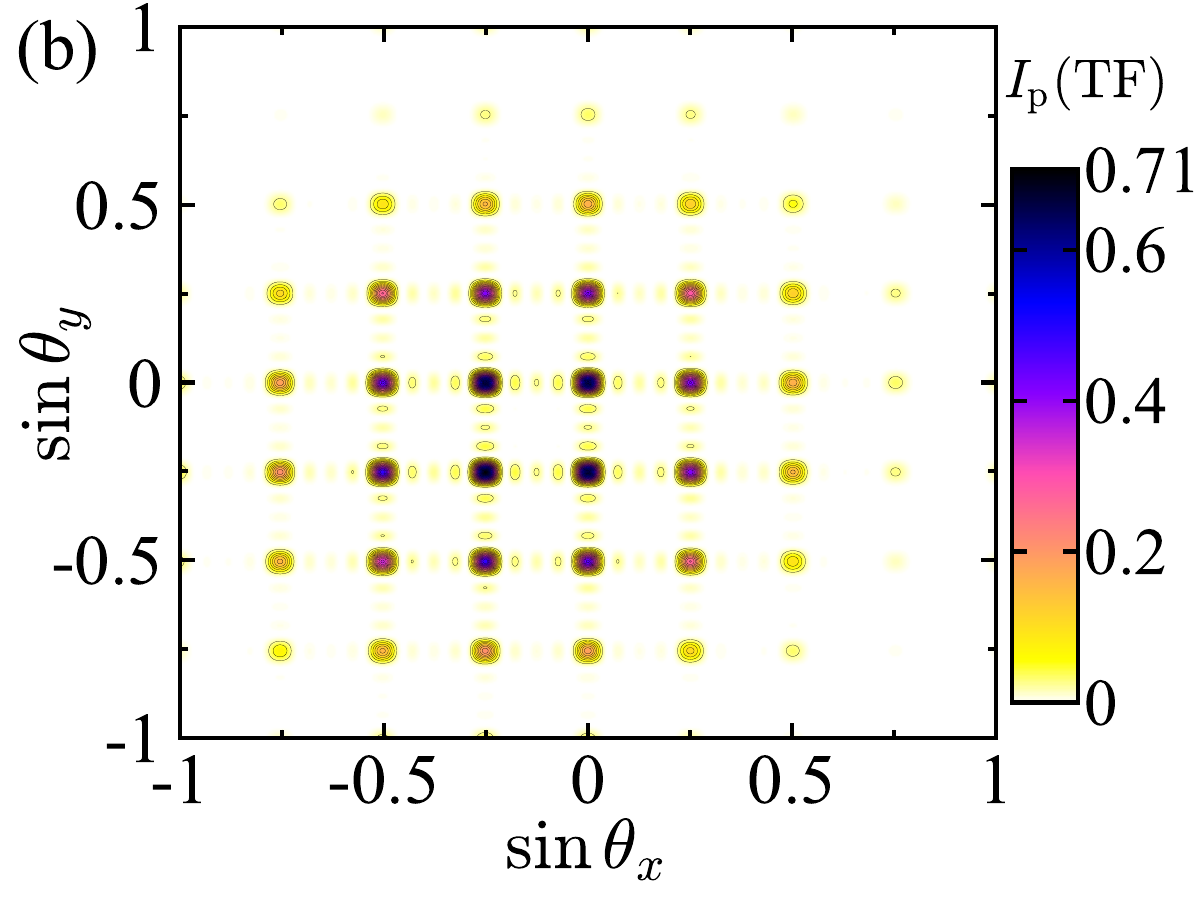} 
    \includegraphics[scale=0.29,trim=0 0 0 0, clip]{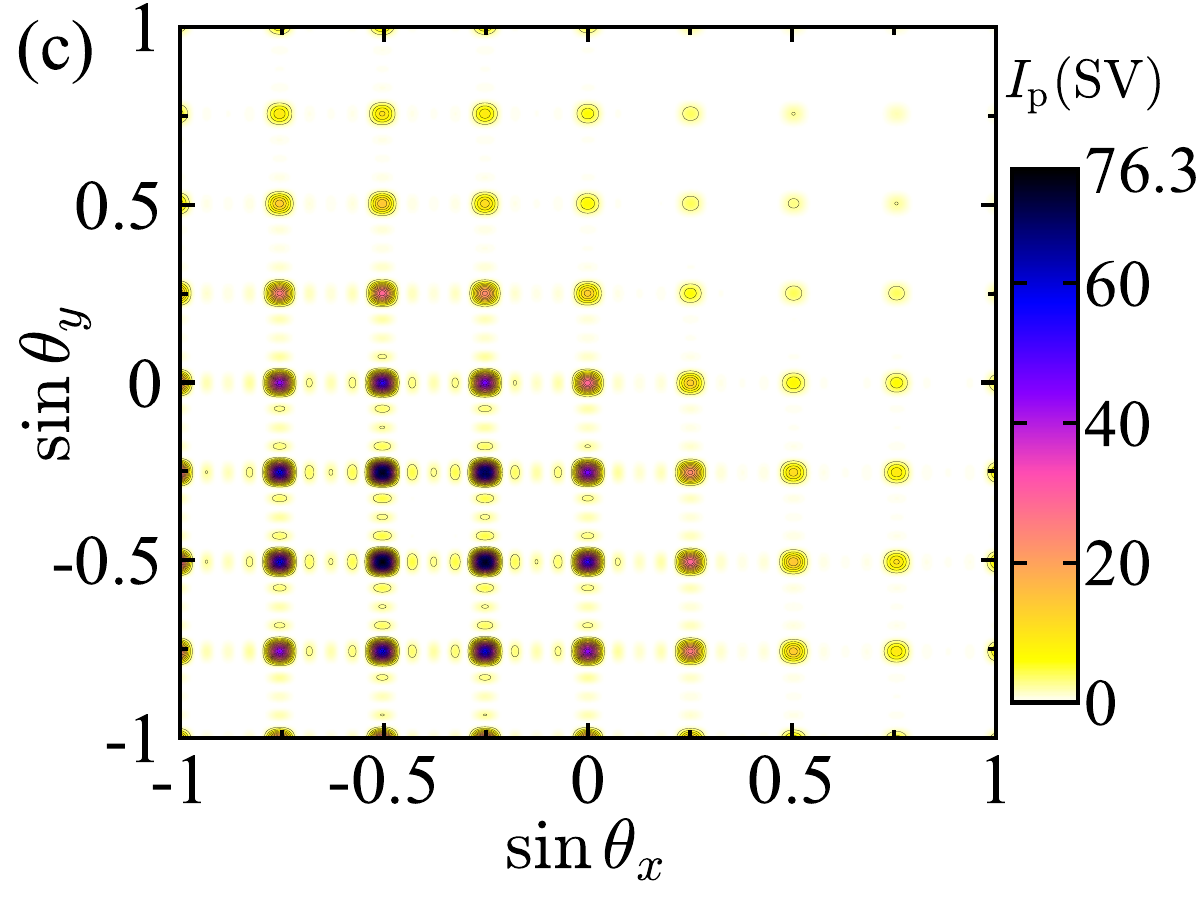}
\caption{
(a), (b), (c) 
Two-dimensional diffraction intensity $I_\text{p}(\theta_x,\theta_y)$ as a function of $\sin(\theta_x)$ and $\sin(\theta_y)$ for NV, TF ($N=0.2$), SV ($N=0.2$) reservoirs, assuming nonzero detuning $\delta/\gamma = 0.2$.
The remaining parameters are set to the same values as those used in Fig.~\ref{fig:5}.
}
\label{fig:6}
\end{figure*}

The role of squeezed-reservoir detuning becomes even more pronounced in Fig.~\ref{fig:6}, where a nonzero detuning $\delta/\gamma = 0.2$ is assumed. It is evident from Figs.~\ref{fig:6}(a) and (b) that the diffraction patterns of the NV and TF reservoirs remain qualitatively similar to those observed at zero detuning [Figs.~\ref{fig:5}(a) and (b)]. In contrast, as shown in Fig.~\ref{fig:6}(c), the SV reservoir undergoes a considerable transformation. Namely, the diffraction intensity becomes highly concentrated into a small subset of diffraction orders, and the dominant peaks are enhanced by several orders of magnitude compared with the undetuned case [Fig.~\ref{fig:5}(c)].
This behavior highlights that the combination of squeezing and finite reservoir detuning provides a powerful mechanism for controlling the angular redistribution of diffracted light. Rather than producing a uniform amplification of all diffraction channels, the detuned squeezed reservoir selectively enhances particular directions in momentum space, thereby enabling highly directional diffraction patterns.

From the above observations, we conclude that the two-dimensional results demonstrate that engineered quantum reservoirs can be employed not only to modify the magnitude of diffraction but also to tailor its spatial distribution and phase structure. Such capabilities may be useful for reservoir-assisted beam steering, diffraction control, programmable optical routing, and spatially structured quantum photonic devices.

\section{Conclusion}\label{sec:conclusions}
 
In this work, we have investigated EIGs in a driven two-level medium coupled to three distinct reservoir types NV, TF, and SV. By combining the reservoir-modified optical susceptibility with Fraunhofer diffraction theory, we analyzed the spatial transmission profile and the resulting one- and two-dimensional diffraction patterns of a weak probe field.
Our investigations demonstrate that reservoir engineering provides a powerful and versatile mechanism for controlling both the magnitude and angular distribution of probe-field diffraction. In the NV case, increasing the control-field decay parameter progressively suppresses transmission maxima and diffraction orders, while larger control-field amplitudes  enhance higher-order channels. Coupling to a thermal reservoir substantially enhances transmission modulation and diffraction intensities, acting as an effective amplifier of the EIG response. We have found that the squeezed-vacuum reservoir generates narrow, high-contrast transmission peaks and selectively amplifies specific diffraction orders. This behavior is a consequence of phase-sensitive reservoir correlations encoded in the squeezing parameter \(M\).

We further showed that the squeezed-reservoir detuning \(\delta/\gamma\) serves as an efficient control parameter for tailoring the angular redistribution of diffracted light. In one-dimensional geometries, the strongest forward diffraction occurs for specific combinations of reservoir photon number and its detuning parameter, where the detuning modifies the relative weights of Fourier components rather than simply scaling overall intensity. In two-dimensional geometries, the SV reservoir produces a significantly larger phase excursion across the transverse plane compared with NV and TF cases. Introducing a finite detuning substantially concentrates diffraction intensity into a small subset of orders, where dominant peaks are several orders of magnitude larger than in the undetuned case. Thus, the combination of squeezing and detuning enables a controlled transition from relatively uniform diffraction patterns to highly directional responses dominated by a few diffraction channels.

These findings establish engineered quantum reservoirs, particularly their detuned squeezed counterparts, as a powerful tool for manipulating EIGs. The demonstrated capabilities including reservoir-assisted beam steering, angular filtering, selective diffraction-order amplification, programmable optical routing, and spatially structured quantum photonic devices, make new ideas for quantum control in driven optical systems.

\section*{Acknowledgments:} H.A.Z. and M. J. acknowledge funding by the Slovak Research and Development Agency under the contract No. APVV-24-0091, and by the grant of The Ministry of Education, Research, Development and Youth of the Slovak Republic under the contract No. VEGA 1/0298/25.
H.A.Z. acknowledges the financial support provided under the postdoctoral fellowship program of P. J. \v{S}af\'{a}rik University in Ko\v{s}ice, Slovakia.

\begin{widetext}
\clearpage
\appendix
\onecolumngrid

\section{}\label{App:A}
The following is a summary of the parameters and reservoir configurations used in this work, along with their definitions.
\begin{table*}[htbp]
	\centering
	\caption{Summary of symbols and parameters.}
	\label{tab:notation}
	\footnotesize
	\begin{ruledtabular}
		\begin{tabular}{lll}
			\toprule
			\textbf{Symbol} & \textbf{Physical meaning} & \textbf{Remarks / definition} \\
			\midrule
			$\omega_0$ & Atomic transition frequency & Transition between $|g\rangle$ and $|e\rangle$. \\
			$\omega_{\text{c}}$ & Control-field frequency & Defines the control detuning $\Delta_{\text{c}}=\omega_{\text{c}}-\omega_0$. \\
			$\omega_{\text{p}}$ & Probe-field frequency & Defines the probe detuning $\Delta_{\text{p}}=\omega_{\text{p}}-\omega_{\text{c}}$. \\
			$\omega_\text{r}$ & Reservoir central frequency & Central frequency of the broadband squeezed reservoir. \\
			\midrule
			$\Delta_{\text{c}}=\omega_{\text{c}}-\omega_0$ & Control-field detuning & Laser detuning from atomic resonance. \\
			$\Delta_{\text{p}}=\omega_{\text{p}}-\omega_{\text{c}}$ & Probe detuning & Horizontal axis in diffraction and contour plots. \\
			$\delta=\omega_\text{r}-\omega_{\text{c}}$ & Reservoir--control detuning & Controls phase beating and non-resonant squeezed effects. \\
			\midrule
			$N$ & Mean photon number of the reservoir & Determines thermal or squeezed reservoir strength. \\
			$M=|M|e^{i\phi}$ & Squeezing correlation parameter & $|M|\le\sqrt{N(N+1)}$, $\phi$ is the squeezing phase. \\
			$\mathcal{R}$ & Reservoir coupling rate & Sets the scale of reservoir-induced decay. \\
			$\gamma$ & Spontaneous emission rate & Vacuum decay contribution. \\
			$\Gamma=\frac{r}{2}(1+2N)$ & Effective decay rate & Modified by thermal or squeezed reservoir occupation. \\
			\midrule
			$\Omega_{\text{c}}(x)=\Omega_{\text{c}0}\sin(k_{\text{c}}x)$ & Control-field Rabi frequency & Position-dependent, peak value $\Omega_{\text{c}0}$. \\
			$\Omega_{\text{p}}$ & Probe-field Rabi frequency & Assumed weak, $\Omega_{\text{p}}\ll\Omega_{\text{c}0}$. \\
			$\eta$ & Control-field decay rate & Governs the exponential envelope $e^{-\eta t}$. \\
			\midrule
			$\chi(\Delta_{\text{p}})$ & Linear susceptibility & Complex quantity: $\chi=\chi'+i\chi''$. \\
			$\chi'(\Delta_{\text{p}})$ & Dispersion (dispersive) component & Real part of susceptibility, $\operatorname{Re}\chi$. \\
			$\chi''(\Delta_{\text{p}})$ & Absorption/gain component & Imaginary part of susceptibility, $\operatorname{Im}\chi$. \\
			\midrule
			$\lambda_{\text{p}}$ & Probe wavelength & Determines diffraction angles. \\
			$\lambda_{\text{c}}$ & Control-field wavelength & Enters the grating period. \\
			$\Lambda_g$ & Grating spatial period & $\Lambda_g = \lambda_{\text{c}}/[2\sin(\theta_c/2)]$. \\
			\midrule
			$\xi$ & Transverse correlation length & $\xi \simeq \sqrt{D\tau_{\text{coh}}}$, width of one grating cell. \\
			$L$ & Medium length & Propagation length along the $z$ direction. \\
			\midrule
			$\theta$ & Diffraction angle (1D) & Used in 1D Fraunhofer diffraction pattern. \\
			$\theta_x,\;\theta_y$ & Diffraction angles (2D) & Used in 2D Fraunhofer diffraction patterns. \\
			\bottomrule
		\end{tabular}
	\end{ruledtabular}
\end{table*}

\newpage

\newpage

\section{}\label{App:B}

The auxiliary response functions $u_{+}^{*}(c)$ and $u_z^{*}(c)$ appearing in Eq.~(\ref{eq:chi_compact}) are obtained from the steady-state solutions of Eqs.~(\ref{eq:bloch_plus}) and (\ref{eq:bloch_z}) in the Laplace domain. Following Ref.~\cite{wu1986parametric}, the explicit forms of these functions for a general squeezed vacuum reservoir are given below. The Laplace variable is $c = i\Delta_{\text{p}}$, and the complex combinations $\kappa = \eta/\gamma + i\delta/\gamma$ and $\kappa^* = \eta/\gamma - i\delta/\gamma$ incorporate the control-field decay rate $\eta$ and the reservoir detuning $\delta/\gamma$.

\begin{equation}
	\begin{aligned}
		u_+^* &= \Biggl\{ 
		\frac{1}{2\mathcal{R}} 
		\Biggl[ \frac{\mathcal{R} + iq}{s + \Gamma - \mathcal{R} + i\delta} + \frac{\mathcal{R} - iq}{s + \Gamma + \mathcal{R} + i\delta} \Biggr] 
		\\[2mm]
		&+ \Omega_{\text{c}}^2 \Biggl[ -\frac{1}{4\mathcal{R}(s + \Gamma - \mathcal{R} + i\delta)} 
		\Biggl( \frac{\mathcal{R} + iq}{\Gamma + \mathcal{R} + \kappa^*} \Bigl[ \frac{\mathcal{R} + iq + 2\eta}{\eta (\eta + \mathcal{R})} + \frac{\gamma M}{\kappa^*(\mathcal{R} + \kappa^*)} \Bigr] 
		+ \frac{\gamma M^*}{\Gamma + \mathcal{R} + \kappa} \Bigl[ \frac{\mathcal{R} + iq + 2\kappa}{\kappa(\mathcal{R} + \kappa)} + \frac{\gamma M}{\eta(\eta + \mathcal{R})} \Bigr] \Biggr)
		\\[1mm]
		&+ \frac{1}{4\mathcal{R}(s + \Gamma + \mathcal{R} + i\delta)} 
		\Biggl( \frac{\mathcal{R} - iq}{\Gamma - \mathcal{R} + \kappa^*} \Bigl[ \frac{\mathcal{R} - iq - 2\eta}{\eta (\eta - \mathcal{R})} + \frac{\gamma M}{\kappa^*(\mathcal{R} - \kappa^*)} \Bigr]
		+ \frac{\gamma M^*}{\Gamma - \mathcal{R} + \kappa} \Bigl[ \frac{\mathcal{R} - iq - 2\kappa}{\kappa(\mathcal{R} - \kappa)} + \frac{\gamma M}{\eta (\eta - \mathcal{R})} \Bigr] \Biggr)
		\\[1mm]
		&+ \frac{1}{4 \mathcal{R} \eta (\mathcal{R} - \eta) (s + \Gamma - \mathcal{R} + 2\eta + i\delta)} 
		\Biggl( \frac{(\mathcal{R} + iq)(\mathcal{R} + iq - 2\eta)}{\Gamma + \mathcal{R} - \kappa} - \frac{\gamma^2 |M|^2}{\Gamma + \mathcal{R} - \kappa^*} \Biggr)
		\\[1mm]
		&+ \frac{1}{4 \mathcal{R} \eta (\mathcal{R} + \eta) (s + \Gamma + \mathcal{R} + 2\eta + i\delta)} 
		\Biggl( \frac{(\mathcal{R} - iq)(\mathcal{R} - iq + 2\eta)}{\Gamma - \mathcal{R} - \kappa} + \frac{\gamma^2 |M|^2}{\Gamma - \mathcal{R} - \kappa^*} \Biggr)
		\\[1mm]
		&\times 2 \Biggl( \frac{\kappa - \Gamma + iq}{(\Gamma - \kappa)^2 - \mathcal{R}^2} + \frac{\gamma M}{(\Gamma - \kappa^*)^2 - \mathcal{R}^2} \Biggr)
		\\[1mm]
		&\times \Biggl[ \frac{\Gamma + \kappa^* - iq}{[(\Gamma + \kappa^*)^2 - \mathcal{R}^2](s + 2\Gamma + \kappa^* + i\delta)} 
		- \frac{\gamma M^*}{[(\Gamma + \kappa)^2 - \mathcal{R}^2](s + 2\Gamma + \kappa + i\delta)}
		\\[1mm]
		&+ \frac{\gamma M}{4 \mathcal{R} \kappa^*} \Bigl( \frac{\mathcal{R} + iq - 2\kappa^*}{(\Gamma + \mathcal{R} - \kappa^*)(\mathcal{R} - \kappa^*)(s + \Gamma + 2\kappa^* - \mathcal{R} + i\delta)} 
		- \frac{\mathcal{R} - iq - 2\kappa^*}{(\Gamma - \mathcal{R} - \kappa^*)(\mathcal{R} + \kappa^*)(s + \Gamma + 2\kappa^* + \mathcal{R} + i\delta)} \Bigr)
		\\[1mm]
		&+ \frac{\gamma M^*}{4 \mathcal{R} \kappa} \Bigl( \frac{\mathcal{R} + iq}{(\Gamma + \mathcal{R} - \kappa)(\mathcal{R} - \kappa)(s + \Gamma + 2\kappa - \mathcal{R} + i\delta)} 
		- \frac{\mathcal{R} - iq}{(\Gamma - \mathcal{R} - \kappa)(\mathcal{R} + \kappa)(s + \Gamma + 2\kappa + \mathcal{R} + i\delta)} \Bigr) \Biggr] \Biggr] 
		\Biggr\}
	\end{aligned}
	\label{eq:C_plus_full}
\end{equation}

\begin{equation}
	\begin{aligned}
		u_z^* &= -2i\Omega_{\text{c}0} \Bigg\{
		\frac{1}{2 \mathcal{R} (s + \Gamma + \mathcal{R} + i\delta)} \Bigg(
		\frac{\mathcal{R} - i q}{\Gamma + \kappa^* - \mathcal{R}}
		- \frac{\gamma M^*}{\Gamma + \kappa - \mathcal{R}}
		\Bigg)
		\\[4pt]
		&\quad
		- \frac{1}{2 \mathcal{R} (s + \Gamma - \mathcal{R} + i\delta)} \Bigg(
		\frac{\mathcal{R} + i q}{\Gamma + \kappa^* + \mathcal{R}}
		+ \frac{\gamma M^*}{\Gamma + \kappa + \mathcal{R}}
		\Bigg)
		\\[4pt]
		&\quad
		- \frac{\Gamma + \kappa^* - i q}{\big[(\Gamma + \kappa^*)^2 - \mathcal{R}^2\big]\, (s + 2\Gamma + \kappa^* + i\delta)}
		+ \frac{\gamma M^*}{\big[(\Gamma + \kappa)^2 - \mathcal{R}^2\big]\, (s + 2\Gamma + \kappa + i\delta)}
		\Bigg\}
	\end{aligned}
	\label{eq:C_z_full}
\end{equation}
In the above expressions we have
$s = c + \eta/\gamma = i\Delta_{\text{p}}/\gamma + \eta/\gamma$ and $q = \Delta_{\text{c}}/\gamma - \delta/\gamma = \dfrac{1}{\gamma}(2\omega_{\text{c}} - \omega_0 - \omega_\text{r})$.

\begin{comment}
	The compact forms of $u_+^*$ and $u_z^*$ are recovered from the full expressions above after simplification:
	
	\begin{align}
		\mathcal{C}_+ &= \frac{1}{\Gamma + i\Delta_{\text{c}} + c + \eta - i\delta}
		\left( \gamma M + \frac{2\Omega_{\text{c}0}^2 \sin^2(k_{\text{c}}x)}{\Gamma + c + \eta + i\delta} \right),
		\\
		\mathcal{C}_z &= \frac{-2i\Omega_{\text{c}0} \sin(k_{\text{c}}x)}{\Gamma + c + \eta + i\delta}
		\left( 1 + \frac{\mathcal{C}_+}{\Gamma + i\Delta_{\text{c}} + c + \eta - i\delta} \right).
	\end{align}
\end{comment}

\end{widetext}

%%%%%%%%%%%%%%%%%%%%%%%%%%%%%%%%%%%%%%%%%%%%%%%%%%%%%%%%%%%%%%%%%%%%%%%%%%%%%%%%%%%%%%%%%%%%%%%%%%%%%

\bibliographystyle{unsrtnat}
\bibliography{bibliography}

@book{scully2012quantum,
author    = {Scully, M. O. and Zubairy, M. S.},
title     = {Quantum Optics},
publisher = {Cambridge University Press},
address   = {Cambridge},
year      = {2012},
doi       = {10.1017/CBO9780511813993}
}

@book{gerry2012introductory,
author    = {Gerry, C. C. and Knight, P. L.},
title     = {Introductory Quantum Optics},
publisher = {Cambridge University Press},
address   = {Cambridge},
year      = {2012},
doi       = {10.1017/CBO9780511791239}
}

@article{stern2013nanoscale,
author  = {Stern, L. and Desiatov, B. and Goykhman, I. and Levy, U.},
title   = {Nanoscale light--matter interactions in atomic cladding waveguides},
journal = {Nat. Commun.},
volume  = {4},
pages   = {1548},
year    = {2013},
doi     = {10.1038/ncomms2554}
}

@article{harris1997eit,
author  = {Harris, S. E.},
title   = {Electromagnetically induced transparency},
journal = {Phys. Today},
volume  = {50},
pages   = {36},
year    = {1997},
doi     = {10.1063/1.881806}
}

@article{fleischhauer2002memory,
author  = {Fleischhauer, M. and Lukin, M. D.},
title   = {Quantum memory for photons: Dark-state polaritons},
journal = {Phys. Rev. A},
volume  = {65},
pages   = {022314},
year    = {2002},
doi     = {10.1103/PhysRevA.65.022314}
}

@article{fleischhauer2005review,
author  = {Fleischhauer, M. and Imamoglu, A. and Marangos, J. P.},
title   = {Electromagnetically induced transparency: Optics in coherent media},
journal = {Rev. Mod. Phys.},
volume  = {77},
pages   = {633--673},
year    = {2005},
doi     = {10.1103/RevModPhys.77.633}
}

@article{mitsunaga1999eig,
author  = {Mitsunaga, M. and Imoto, N.},
title   = {Observation of an electromagnetically induced grating in cold sodium atoms},
journal = {Phys. Rev. A},
volume  = {59},
pages   = {4773--4776},
year    = {1999},
doi     = {10.1103/PhysRevA.59.4773}
}

@article{deng2003ultraslow,
author  = {Deng, L. and Payne, M. G.},
title   = {Inhibiting the onset of the three-photon destructive interference in ultraslow propagation-enhanced four-wave mixing with dual induced transparency},
journal = {Phys. Rev. Lett.},
volume  = {91},
pages   = {243902},
year    = {2003},
doi     = {10.1103/PhysRevLett.91.243902}
}

@article{wen2011talbot,
author  = {Wen, J. and Du, S. and Chen, H. and Xiao, M.},
title   = {Electromagnetically induced Talbot effect},
journal = {Appl. Phys. Lett.},
volume  = {98},
pages   = {081108},
year    = {2011},
doi     = {10.1063/1.3559610}
}

@article{ba2012magnetic,
author  = {Ba, N. and Wu, X.-Y. and Liu, X.-J. and Zhang, S.-Q. and Wang, J.},
title   = {Electromagnetically induced grating in an atomic system with a static magnetic field},
journal = {Opt. Commun.},
volume  = {285},
pages   = {3792--3797},
year    = {2012},
doi     = {10.1016/j.optcom.2012.05.015}
}

@article{brown2005switching,
author  = {Brown, A. W. and Xiao, M.},
title   = {All-optical switching and routing based on an electromagnetically induced absorption grating},
journal = {Opt. Lett.},
volume  = {30},
pages   = {699--701},
year    = {2005},
doi     = {10.1364/OL.30.000699}
}

@article{zhai2001bistability,
author  = {Zhai, P.-W. and Su, X.-M. and Gao, J.-Y.},
title   = {Optical bistability in electromagnetically induced grating},
journal = {Phys. Lett. A},
volume  = {289},
pages   = {27--32},
year    = {2001},
doi     = {10.1016/S0375-9601(01)00576-X}
}

@article{chen2017router,
author  = {Chen, Y.-Y. and Liu, Z.-Z. and Wan, R.-G.},
title   = {Beam splitter and router via an incoherent pump-assisted electromagnetically induced blazed grating},
journal = {Appl. Opt.},
volume  = {56},
pages   = {5736--5742},
year    = {2017},
doi     = {10.1364/AO.56.005736}
}

@article{yan2001switching,
author  = {Yan, M. and Rickey, E. G. and Zhu, Y.},
title   = {Observation of absorptive photon switching by quantum interference},
journal = {Phys. Rev. A},
volume  = {64},
pages   = {041801},
year    = {2001},
doi     = {10.1103/PhysRevA.64.041801}
}

@article{zhang2018talbot,
author  = {Zhang, Z. and Liu, X. and Zhang, D. and Sheng, J. and Zhang, Y. and Zhang, Y. and Xiao, M.},
title   = {Observation of electromagnetically induced Talbot effect in an atomic system},
journal = {Phys. Rev. A},
volume  = {97},
pages   = {013603},
year    = {2018},
doi     = {10.1103/PhysRevA.97.013603}
}

@article{deng2001rubidium,
author  = {Deng, L. and Hagley, E. W. and Payne, M.},
title   = {Electromagnetically induced gratings in rubidium vapor},
journal = {Phys. Rev. Lett.},
volume  = {86},
pages   = {5409--5412},
year    = {2001},
doi     = {10.1103/PhysRevLett.86.5409}
}

@article{kang2003efficient,
author  = {Kang, H. and Hernandez, G. and Zhu, Y.},
title   = {Electromagnetically induced grating: Efficient diffraction of weak probe light in a rubidium vapor},
journal = {Phys. Rev. A},
volume  = {67},
pages   = {053824},
year    = {2003},
doi     = {10.1103/PhysRevA.67.053824}
}

@article{yan2003highorder,
author  = {Yan, M. and Kang, H. and Zhu, Y.},
title   = {High-order diffraction in electromagnetically induced gratings with rubidium vapor},
journal = {Phys. Rev. A},
volume  = {68},
pages   = {023812},
year    = {2003},
doi     = {10.1103/PhysRevA.68.023812}
}

@article{cheng2016JPhysB,
author  = {Cheng, G.-L. and Cong, L. and Chen, A.-X.},
title   = {Two-dimensional electromagnetically induced grating via gain and phase modulation in a two-level system},
journal = {J. Phys. B: At. Mol. Opt. Phys.},
volume  = {49},
number  = {8},
pages   = {085501},
year    = {2016},
doi     = {10.1088/0953-4075/49/8/085501}
}

@article{wang2022rydberg,
author  = {Wang, B. and Yan, D. and Liu, Y. and Wu, J.},
title   = {Two-dimension asymmetric electromagnetically induced grating in Rydberg atoms},
journal = {Photonics},
volume  = {9},
number  = {10},
pages   = {674},
year    = {2022},
doi     = {10.3390/photonics9100674}
}

@article{asadpour2021azimuthal,
author  = {Asadpour, S. H. and Kirova, T. and Qian, J. and Hamedi, H. R. and Juzeli={u}nas, G. and Paspalakis, E.},
title   = {Azimuthal modulation of electromagnetically induced grating using structured light},
journal = {Sci. Rep.},
volume  = {11},
pages   = {20721},
year    = {2021},
doi     = {10.1038/s41598-021-00141-9}
}

@book{walls2008quantum,
author    = {Walls, D. F. and Milburn, G. J.},
title     = {Quantum Optics},
edition   = {2},
publisher = {Springer},
address   = {Berlin},
year      = {2008},
doi       = {10.1007/978-3-540-28574-8}
}

@article{xu2019sensing,
author  = {Xu, C. and Zhang, L. and Huang, S. and Ma, T. and Xiao, M.},
title   = {Sensing and tracking enhanced by quantum squeezing},
journal = {Photon. Res.},
volume  = {7},
pages   = {A14--A19},
year    = {2019},
doi     = {10.1364/PRJ.7.000A14}
}

@article{grace2020gyroscope,
author  = {Grace, M. R. and Gagatsos, C. N. and Zhuang, Q. and Guha, S.},
title   = {Quantum-enhanced fiber-optic gyroscopes using quadrature squeezing and continuous-variable entanglement},
journal = {Phys. Rev. Applied},
volume  = {14},
pages   = {034065},
year    = {2020},
doi     = {10.1103/PhysRevApplied.14.034065}
}

@article{malitesta2023distributed,
author  = {Malitesta, M. and Smerzi, A. and Pezz`e, L.},
title   = {Distributed quantum sensing with squeezed-vacuum light in a configurable array of Mach--Zehnder interferometers},
journal = {Phys. Rev. A},
volume  = {108},
pages   = {032621},
year    = {2023},
doi     = {10.1103/PhysRevA.108.032621}
}

@article{zhang2022quadratic,
author  = {Zhang, S.-D. and "Ozdemir, \c{S}. K. and Qiu, C.-W. and Nori, F. and Jing, H.},
title   = {Squeezing-enhanced quantum sensing with quadratic optomechanics},
journal = {Optica},
volume  = {9},
pages   = {345--352},
year    = {2022},
doi     = {10.1364/OPTICA.9.000345}
}

@article{caves1981noise,
author  = {Caves, C. M.},
title   = {Quantum-mechanical noise in an interferometer},
journal = {Phys. Rev. D},
volume  = {23},
pages   = {1693--1708},
year    = {1981},
doi     = {10.1103/PhysRevD.23.1693}
}

@article{vahlbruch2016fifteen,
author  = {Vahlbruch, H. and Mehmet, M. and Danzmann, K. and Schnabel, R.},
title   = {Detection of 15 dB squeezed states of light and their application for the absolute calibration of photoelectric quantum efficiency},
journal = {Phys. Rev. Lett.},
volume  = {117},
pages   = {110801},
year    = {2016},
doi     = {10.1103/PhysRevLett.117.110801}
}

@article{braunstein2005continuous,
author  = {Braunstein, S. L. and van Loock, P.},
title   = {Quantum information with continuous variables},
journal = {Rev. Mod. Phys.},
volume  = {77},
pages   = {513--577},
year    = {2005},
doi     = {10.1103/RevModPhys.77.513}
}

@article{gisin2007communication,
author  = {Gisin, N. and Thew, R.},
title   = {Quantum communication},
journal = {Nat. Photonics},
volume  = {1},
pages   = {165--171},
year    = {2007},
doi     = {10.1038/nphoton.2007.22}
}

@article{schnabel2017squeezed,
author  = {Schnabel, R.},
title   = {Squeezed states of light and their applications in laser interferometers},
journal = {Phys. Rep.},
volume  = {684},
pages   = {1--51},
year    = {2017},
doi     = {10.1016/j.physrep.2017.04.001}
}

@book{loudon2000light,
author    = {Loudon, R.},
title     = {The Quantum Theory of Light},
edition   = {3},
publisher = {Oxford University Press},
address   = {Oxford},
year      = {2000},
doi       = {10.1093/acprof:oso/9780198501763.001.0001}
}

@book{walls2008quantum2,
author    = {Walls, D. F. and Milburn, G. J.},
title     = {Quantum Optics},
edition   = {2},
publisher = {Springer},
address   = {Berlin},
year      = {2008},
doi       = {10.1007/978-3-540-28574-8}
}

@article{wu1986parametric,
author  = {Wu, L.-A. and Kimble, H. J. and Hall, J. L. and Wu, H.},
title   = {Generation of squeezed states by parametric down conversion},
journal = {Phys. Rev. Lett.},
volume  = {57},
pages   = {2520--2523},
year    = {1986},
doi     = {10.1103/PhysRevLett.57.2520}
}

@article{anderson1995quadrature,
author  = {Anderson, M. E. and Beck, M. and Raymer, M. G. and Bierlein, J. D.},
title   = {Quadrature squeezing with ultrashort-pulse optical parametric oscillators},
journal = {Opt. Lett.},
volume  = {20},
pages   = {620--622},
year    = {1995},
doi     = {10.1364/OL.20.000620}
}

@article{silberhorn2001epr,
author  = {Silberhorn, C. and Lam, P. K. and Wei{\ss}, O. and K{"o}nig, F. and Korolkova, N. and Leuchs, G.},
title   = {Generation of continuous variable Einstein--Podolsky--Rosen entanglement via the Kerr nonlinearity in an optical fiber},
journal = {Phys. Rev. Lett.},
volume  = {86},
pages   = {4267--4270},
year    = {2001},
doi     = {10.1103/PhysRevLett.86.4267}
}

@article{caves1985formalism1,
author  = {Caves, C. M. and Schumaker, B. L.},
title   = {New formalism for two-photon quantum optics. I. Quadrature phases and squeezed states},
journal = {Phys. Rev. A},
volume  = {31},
pages   = {3068--3092},
year    = {1985},
doi     = {10.1103/PhysRevA.31.3068}
}

@article{schumaker1985formalism2,
  author  = {Schumaker, B. L. and Caves, C. M.},
  title   = {New formalism for two-photon quantum optics. II. Mathematical foundation and compact notation},
  journal = {Phys. Rev. A},
  volume  = {31},
  pages   = {3093--3111},
  year    = {1985},
  doi     = {10.1103/PhysRevA.31.3093}
}

@article{lu1998interference,
  author  = {Lu, Y. J. and Ou, Z. Y.},
  title   = {Observation of nonclassical quantum interference between independent squeezed vacua},
  journal = {Phys. Rev. Lett.},
  volume  = {80},
  pages   = {1857--1860},
  year    = {1998},
  doi     = {10.1103/PhysRevLett.80.1857}
}

@article{christ2011multimode,
  author  = {Christ, A. and Laiho, K. and Eckstein, A. and Cassemiro, K. N. and Silberhorn, C.},
  title   = {Probing multimode squeezing with correlation functions},
  journal = {New J. Phys.},
  volume  = {13},
  pages   = {033027},
  year    = {2011},
  doi     = {10.1088/1367-2630/13/3/033027}
}

@article{roslund2014quantum,
  author  = {Roslund, J. and de Ara\'ujo, R. M. and Jiang, S. and Fabre, C. and Treps, N.},
  title   = {Wavelength-multiplexed quantum networks with ultrafast frequency combs},
  journal = {Nat. Photonics},
  volume  = {8},
  pages   = {109--112},
  year    = {2014},
  doi     = {10.1038/nphoton.2013.340}
}

@book{gardiner2004noise,
  author    = {Gardiner, C. W. and Zoller, P.},
  title     = {Quantum Noise: A Handbook of Markovian and Non-Markovian Quantum Stochastic Methods with Applications to Quantum Optics},
  publisher = {Springer},
  address   = {Berlin},
  year      = {2004},
  doi       = {10.1007/978-3-662-05328-3}
}

@article{collett1984squeezing,
  author  = {Collett, M. J. and Gardiner, C. W.},
  title   = {Squeezing of intracavity and traveling-wave light fields produced in parametric amplification},
  journal = {Phys. Rev. A},
  volume  = {30},
  pages   = {1386--1391},
  year    = {1984},
  doi     = {10.1103/PhysRevA.30.1386}
}

@article{gardiner1986inhibition,
  author  = {Gardiner, C. W.},
  title   = {Inhibition of atomic phase decays by squeezed light: A direct effect of squeezing},
  journal = {Phys. Rev. Lett.},
  volume  = {56},
  pages   = {1917--1920},
  year    = {1986},
  doi     = {10.1103/PhysRevLett.56.1917}
}

@article{serikawa2017broadband,
  author  = {Serikawa, T. and Yoshikawa, J. and Makino, K. and Furusawa, A.},
  title   = {Creation and measurement of broadband squeezed vacuum from a ring optical parametric oscillator},
  journal = {Opt. Express},
  volume  = {25},
  pages   = {28357},
  year    = {2017},
  doi     = {10.1364/OE.25.028357}
}

@article{chelkowski2024coherent,
  author  = {Chelkowski, S. and Vahlbruch, H. and Danzmann, K. and Schnabel, R.},
  title   = {Coherent control of broadband vacuum squeezing},
  journal = {Phys. Rev. A},
  volume  = {75},
  pages   = {043814},
  year    = {2007},
  doi     = {10.1103/PhysRevA.75.043814}
}

@article{polzik1992spectroscopy,
  author  = {Polzik, E. S. and Carri, J. and Kimble, H. J.},
  title   = {Spectroscopy with squeezed light},
  journal = {Phys. Rev. Lett.},
  volume  = {68},
  pages   = {3020--3023},
  year    = {1992},
  doi     = {10.1103/PhysRevLett.68.3020}
}

@article{zhu1990fluorescence,
  author  = {Zhu, Y. and Dey, T. N.},
  title   = {Modified resonance fluorescence of a two-level atom in a squeezed vacuum},
  journal = {Phys. Rev. A},
  volume  = {41},
  pages   = {254--263},
  year    = {1990},
  doi     = {10.1103/PhysRevA.41.254}
}

@article{jin1990bandwidth,
  author  = {Jin, R.-B. and Tang, S.-H.},
  title   = {Resonance fluorescence spectrum of a two-level atom in a finite-bandwidth squeezed vacuum},
  journal = {Phys. Rev. A},
  volume  = {42},
  pages   = {3277--3285},
  year    = {1990},
  doi     = {10.1103/PhysRevA.42.3277}
}

@article{hassan2013nonlinear,
  author  = {Hassan, S. S. and Alharbey, R. A.},
  title   = {Nonlinear optical processes in multi-level atomic systems},
  journal = {Nonlinear Optics and Quantum Optics},
  volume  = {45},
  pages   = {171--186},
  year    = {2013}
}

@article{lu1998frequency,
  author  = {Lu, N.},
  title   = {Frequency-dependent effects of squeezed vacuum on resonance fluorescence},
  journal = {Phys. Rev. A},
  volume  = {57},
  pages   = {2445--2452},
  year    = {1998},
  doi     = {10.1103/PhysRevA.57.2445}
}

@article{parkins1993twophoton,
  author  = {Parkins, A. S. and Marte, P. and Zoller, P. and Carmichael, H. J.},
  title   = {Squeezed-light-induced two-photon transitions in atom--cavity systems},
  journal = {Phys. Rev. A},
  volume  = {48},
  pages   = {758--765},
  year    = {1993},
  doi     = {10.1103/PhysRevA.48.758}
}

@article{kaertner1990squeezed,
  author  = {Kaertner, F. X. and Russer, P.},
  title   = {Generation of squeezed microwave states by a dc-pumped degenerate parametric Josephson junction oscillator},
  journal = {Phys. Rev. A},
  volume  = {42},
  pages   = {5601--5612},
  year    = {1990},
  doi     = {10.1103/PhysRevA.42.5601}
}

@phdthesis{qiu2022jtwpa,
  author  = {Qiu, J.},
  title   = {Broadband Squeezed Microwave States in Josephson Traveling-Wave Parametric Amplifiers},
  school  = {Massachusetts Institute of Technology},
  year    = {2022},
  url     = {https://dspace.mit.edu/handle/1721.1/150037}
}

@inproceedings{haider2019jtwpa,
  author  = {Haider, M. and Russer, J. A. and Patino, J. A. and Jirauschek, C. and Russer, P.},
  title   = {A Josephson traveling wave parametric amplifier for quantum coherent signal processing},
  booktitle = {2019 IEEE MTT-S International Microwave Symposium (IMS)},
  pages   = {956--958},
  year    = {2019},
  doi     = {10.1109/MWSYM.2019.8700884}
}

@article{corcoles2011protecting,
  author  = {Corcoles, A. D. and Chow, J. M. and Gambetta, J. M. and Rigetti, C. and Rozen, J. R. and Keefe, G. A. and Ketchen, M. B. and Steffen, M.},
  title   = {Protecting superconducting qubits from external sources of loss and heat},
  journal = {Appl. Phys. Lett.},
  volume  = {99},
  pages   = {181906},
  year    = {2011},
  doi     = {10.1063/1.3658630}
}

@article{krinner2019cryogenic,
  author  = {Krinner, S. and Storz, S. and Kurpiers, P. and Magnard, P. and Heinsoo, J. and Keller, R. and Lütolf, J. and Eichler, C. and Wallraff, A.},
  title   = {Engineering cryogenic setups for 100-qubit scale superconducting circuit systems},
  journal = {EPJ Quantum Technol.},
  volume  = {6},
  pages   = {2},
  year    = {2019},
  doi     = {10.1140/epjqt/s40507-019-0072-3}
}

@article{epjqt2025cryogenic,
  author  = {Smith, T. and others},
  title   = {Cryogenic thermal modeling of microwave high density signaling},
  journal = {EPJ Quantum Technol.},
  volume  = {12},
  pages   = {124},
  year    = {2025},
  doi     = {10.1140/epjqt/s40507-025-00427-1}
}

\end{document}